\documentclass[12pt,preprint]{aastex}
\usepackage{amsmath}

\shorttitle{Neutrino Cooled Accretion Disk and Its Stability}
\begin{document}

\title{Neutrino Cooled Accretion Disk and Its Stability}
\author{N. Kawanaka\altaffilmark{1} and S. Mineshige\altaffilmark{1}}
\altaffiltext{1}{Yukawa Institute for Theoretical Physics, Kyoto University,
Kyoto 606-8502, Japan}
\email{norita@yukawa.kyoto-u.ac.jp}

\begin{abstract}
We investigate the structure and stability of hypercritical accretion flows around stellar-mass black holes,
 taking into account neutrino cooling, lepton conservation, and firstly
 a realistic equation of state in order to properly treat the dissociation of nuclei.  We obtain the radial distributions
 of physical properties, such as density, temperature and electron fraction, for various mass accretion rates $0.1\sim 10M_{\odot }{\rm s}^{-1}$.
  We find that, depending on mass accretion rates, different physics affect considerably the structure of the disk; most important physics is
 (1) the photodissociation of nuclei around $r\sim 100r_g$ for relatively low mass accretion rates ($\dot{M}\sim 0.01-0.1M_{\odot}~{\rm s}^{-1}$),
 (2) efficient neutrino cooling around $r\sim 10-100r_g$ for moderately high mass accretion rate ($\dot{M}\sim 0.2-1.0M_{\odot}{\rm s}^{-1}$),
 and (3) neutrino trapping ($r\sim 3-10r_g$) for very high mass accretion rate ($\dot{M}\gtrsim 2.0M_{\odot}{\rm s}^{-1}$).
  We also investigate the stability of hypercritical accretion flows by drawing the thermal equilibrium curves, and find that efficient neutrino cooling
 makes the accretion flows rather stable against both thermal and viscous modes.
\end{abstract}

\section{INTRODUCTION}
Gamma-Ray Bursts (GRBs) are the most energetic explosions in Universe, which are believed to release energies up to $10^{51-53}$ergs in only a few
 tens of seconds.  Their observational behavior (lightcurves of
 prompt emission and afterglow, spectrum, etc.) are well explained by the relativistic fireball shock model
 (for reviews, see e.g. Piran 1999, and M\'{e}sz\'{a}ros 2002).   The central engine which produces hot and baryon-poor plasma
 (i.e. a fireball) is, however, totally hidden from direct observation, and so the engine and fireball-making processes have not well been
 understood yet.  Nevertheless, there is one popular model for the energy sources of GRBs which has been intensely studied by now---a hyperaccreting
 black hole model (Narayan et al. 1992, 2001).  In this model, the energy of relativistic jets that make intense gamma ray emissions is produced via the accretion of a massive
 disk ($\sim 0.1M_{\odot }$) onto a stellar mass black hole.  Such systems are expected after several energetic phenomena; mergers of double neutron star
 binaries (Eichler et al. 1989), neutron star/black hole binaries (Paczy\'{n}ski 1991), white dwarf/black hole binaries (Fryer et al. 1999), black hole/He
 star binaries (Fryer and Woosley 1998) and failed supernovae (or collapsars, Woosley 1993, Paczy\'{n}ski 1998, MacFadyen and Woosley 1999).

  There are several reasons why the hyperaccreting black hole model is widely accepted.  First, this model can explain both the variability timescale
 ($\delta t \sim 1{\rm msec}$) and the duration timescale ($T_{\rm dur}\sim 1-100{\rm sec}$); the former timescale represents the compactness of the central
 engine--the inner radius of the disk is smaller than $c\delta t\sim 3\times 10^7$ cm---and the latter represents the accretion timescale---the outer radius
 of the disk $r_{\rm out}$ can be estimated by $T_{\rm dur}\sim r_{\rm out}/(\alpha c_s(H/r_{\rm out}))=1/(\alpha \Omega (H/r_{\rm out})^2)$,
 where $\alpha$, $H$ and $c_s$ are the viscosity parameter (see below), scale height of the disk and sound velocity, respectively.  Second, the relativistic jets which are believed to
 be produced in GRBs formed in accretion disk systems, as can be seen in active galactic nuclei or microquasars.

In such a massive accretion disk, photons are almost trapped because the flow become highly optically thick, and so radiative cooling is not efficient.
  However, as the flow become dense ($\sim 10^7{\rm g}~{\rm cm}^{-3}$)and hot ($\sim 10^{10}{\rm K}$), thermal neutrino emissions will set in, and
 become the most efficient cooling process.  Due to this nature, such disks (or flows) are called \lq\lq neutrino-dominated accretion flows \rq\rq (NDAFs).
 The NDAF model can give explanations for not only the energetics of GRBs but also the processes of making the relativistic and baryon-poor fireballs
 through the energy release by neutrino annihilation above the disk (Popham et al. 1999).  In order to predict the amount of energy deposited through
 this process precisely, it is necessary to solve the disk structure taking into account various kinds of microphysics because when calculating total
 luminosity of neutrino annihilation we should know where and how much neutrinos are emitted from the disk.
  First, in the region with a moderate density and temperature, the fraction of heavy nuclei (e.g.
 $^{56}{\rm Fe}$, $^{62}{\rm Ni}$, etc.) can be high.  However, around $r\sim 100r_g$ these nuclei are photodissociated because the temperature of the disk
 rises.  This process consumes the internal energy of the accretion flow and so effectively works as cooling.
  Secondary, in such a high
 density disk electrons are highly degenerated and this will affect the equation of state (EoS), because electron degeneracy pressure will contribute
 to total pressure, and neutrino emission, because degenerate electrons tend to be captured by protons and electron neutrino emissivity becomes larger.
  Especially, by the latter effect the accretion flow are supposed to be highly neutron-rich.  Finally, the accretion flow can be optically thick
 even for neutrinos in the innermost region, so the energy and number density of neutrinos will be advected with accreting matter.  All of
 these effects should be taken into account when we analyze the structure of the disk and its stability.
 
There are many previous studies on NDAFs.  Popham et al. (1999) performed relativistic calculations for the disk structure in Kerr spacetime
 with neutrino-thin approximation.  Narayan et al. (2001) extended their discussion to the low luminosity regime in the context of CDAF
 (convection-dominated accretion flow), and Kohri \& Mineshige (2002) studied the effects of highly degenerated electrons and nucleons on neutrino
 emissions and the pressure in the disk.  Di Matteo et al. (2002) and Kohri et al. (2005) considered the neutrino trapping effect, which is not
 negligible when mass accretion rate is very high.  Gu et al. (2006) discussed the contribution from the neutrino-thick region to neutrino emissions
 and showed that sufficiently large amount of energy for GRBs would be released by neutrino pair annihilation.  Recently, Chen \& Beloborodov (2006)
 calculated the structure of hyper-critical accretion flows around Kerr black holes, fully taking into account the effect of general relativity.
  In their calculations, however, the detailed physics shown above are not fully considered but only part of them approximately.
  In this paper, we take into account the effects of neutrino emission and trapping in NDAFs, and employ the realistic EOS
 in order to see the nuclear composition and dissociation in the outer region of the flow consistently.

The plan of this paper is as follows.  In Section 2, we show the fundamental equations and assumptions describing the stational behavior of NDAFs.
  In Section 3, we show the calculational results and implications of them.  Finally in Section 4. we discuss our results and compare them with previous
 studies. 

\section{MODELS}
In this section, we present the basic equations and assumptions that describe the situations of our interest.
  We assume that the accretion flow we are considering is steady and
  axisymmetric.  In addition, we assume that the gravity is described with pseudo-Newtonian potential (Paczy\'{n}ski and Wiita, 1980), i.e.
\begin{eqnarray}
\Psi _{\rm PN}=-\frac{GM}{r-r_g},
\end{eqnarray}
and the angular velocity and specific angular momentum of gas
particles in circular motion are assumed to be Keplerian, i.e.
\begin{eqnarray}
\Omega(r)&=&\sqrt{\frac{GM}{(r-r_g)^2 r}}, \\
l(r)&=&r^2 \Omega(r),
\end{eqnarray}
The expressions for mass conservation, angular momentum conservation, energy conservation, lepton fraction conservation, hydrostatic balance and
 $\alpha$ viscosity are, respectively,
\begin{eqnarray}
-2\pi r \Sigma v_r &=& \dot{M}, \\
-r^3 \nu \Sigma \frac{d\Omega(r)}{dr}&=&\frac{\dot{M}}{2\pi} (l(r)-l(r_{\rm in})), \\
\frac{\Sigma}{m_p}v_r \left( \frac{de}{dr}-\frac{m_p p}{\rho ^2}\frac{d\rho}{dr} \right) &=& Q^{+}-Q^{-}, \\
\frac{\Sigma}{m_p} v_r \frac{dY_l}{dr}&=&2\left(F_{n, \bar{\nu}_e}-F_{n, \nu_e}\right), \\
\frac{p}{\rho}&=&\Omega(r)^2 H^2, \\
\rho \nu r \frac{d\Omega(r)}{dr}&=&-\alpha p, 
\end{eqnarray}
where $r$, $\Sigma$, $l(r)$, $\rho$, $H$, $v_r$, $\nu$, $\Omega$, $p$, $e$ and $Y_l$ denote the radius, surface mass density,
 specific angular momentum at the radius $r$, baryon mass density, scale height, radial velocity, kinematic viscosity coefficient,
 angular velocity, pressure, internal energy per baryon, and lepton fraction, respectively; $\dot{M}$, $\alpha$, $m_p$ are the mass accretion rate,
 viscosity parameter and mass of a proton, respectively; $Q^+$, $Q^-$, $F_{n, \nu_e}$ and $F_{n, \bar{\nu}_e}$ are
 the viscous heating rate per unit area, cooling rate per unit area,
 neutrino number flux, anti-neutrino number flux, respectively.  We
 set the inner radius of the accretion disk, $r_{\rm in}$, as the radius of the
 innermost stable circular orbit of the central black hole; $r_{\rm in}=3r_g$.
  The surface mass density is defined as
\begin{eqnarray}
\Sigma=2\rho H, 
\end{eqnarray}
and the dissipated energy per unit area is
\begin{eqnarray}
Q^+=\nu \Sigma \left( r \frac{d\Omega(r)}{dr} \right)^2,
\end{eqnarray}
using the above expression of $\Omega(r)$.  On the other hand, the
cooling rate per unit area is
\begin{eqnarray}
Q^-=\sum_{\nu_i} Q^-_{\nu_i},
\end{eqnarray}
where $Q^-_{\nu_i}$ is the cooling rate per unit area via $i$-neutrino
emission, each of which is calculated below.
 
In calculating the neutrino energy flux and the neutrino number flux, we take into account the various kinds of
 reaction; electron-positron capture onto nucleons/nuclei, electron-positron pair annihilation,
 and plasmon decay.  The emissivities and reaction rates of these processes are calculated based on the formulae given by
 Bruenn (1985), Ruffert et al. (1996), and Burrows \& Thompson (2002).   In order to solve the energy equation, the lepton number equation, and the hydrostatic
 equilibrium equation properly, we should know the neutrino energy flux and number flux on the disk surface and the neutrino energy density and
 number density in the disk.  We evaluate these physical values of neutrinos by adopting the approximation by Popham \& Narayan (1995)
 and Di Matteo et al. (2002).  According to their procedure, the neutrino energy flux, energy density, number flux and number density are given by
\begin{eqnarray}
Q^- _{\nu_i}&=&2\cdot \frac{(c/4) u_{\nu_i,0}}{(3/4)\left[\tau_{E,\nu_i}/2+1/\sqrt{3}+1/(3\tau_{E,a,\nu_i})\right]}, \\
u_{\nu_i}&=&\frac{u_{\nu_i,0}\left( 3\tau_{E,\nu_i}+2\sqrt{3} \right)}{3\tau_{E,\nu_i}+2\sqrt{3}+2/\tau_{E,a,\nu_i}}, \\
F_{n,\nu_i}&=&\frac{(c/4) n_{\nu_i,0}}{(3/4)\left[\tau_{n,\nu_i}/2+1/\sqrt{3}+1/(3\tau_{n,a,\nu_i})\right]}, \\
n_{\nu_i}&=&\frac{n_{\nu_i,0}\left( 3\tau_{n,\nu_i}+2\sqrt{3} \right)}{3\tau_{n,\nu_i}+2\sqrt{3}+2/\tau_{n,a,\nu_i}},
\end{eqnarray}
respectively.  Here, $\tau_{E,a,\nu_i}$ and $\tau_{n,a,\nu_i}$ are the spectral-averaged neutrino-absorption optical depth for
 energy transport and number transport, respectively, and $\tau_{E,\nu_i}$ and $\tau_{n,\nu_i}$ are
 the total neutrino optical depth (including scattering) for energy transport and number transport, respectively.
  In the above expressions, $u_{\nu_i,0}$ and $n_{\nu_i,0}$ are defined as
\begin{eqnarray}
u_{\nu_i,0}&=&\frac{k_B^4 T^4}{2\pi^2\hbar ^3 c^3}F_3 \left(\mu_{\nu_i}/k_B T \right), \\
n_{\nu_i,0}&=&\frac{k_B^3 T^3}{2\pi^2\hbar ^3 c^3}F_2 \left(\mu_{\nu_i}/k_B T \right),
\end{eqnarray}
where $F_m(\eta )$ is the standard Fermi-Dirac integral given by
\begin{eqnarray}
F_m(\eta)=\int_0^{\infty}dx \frac{x^m}{e^{x-\eta}+1},
\end{eqnarray}
and $\mu_{\nu_i}$ represents the effective neutrino chemical potentials.  The calculation procedures of neutrino opacities
 and effective neutrino chemical potentials are given in Appendix A.

Most of the previous studies on the structures of NDAFs, a simplified equation of state and nuclear compositions (i.e. all nuclei
 which exist in the flow are only $\alpha$ particles) were assumed and so the estimation of the photodissociation-cooling effects and 
 the value of electron fraction $Y_e$ might not be accurate in the computational results.  In this study, we employ a realistic equation of
 state by Lattimer \& Swesty (1991), which is widely used in supernova simulations.  By using this, we successfully take into account the effects
 of heavy nuclei-dissociation in the calculation.

Finally, we mention the outer boundary conditions that we impose in computing the structures of NDAF.  We assume that the inflowing
 gas is composed primarily of neutron-rich iron group nuclei (i.e. $^{56}{\rm Fe}$, $^{58}{\rm Fe}$, $^{62}{\rm Ni}$ and so on), and the
 electron fraction $Y_e$ is $\approx 0.42$.  Such compositions are often realized in the detailed calculations of the core of a
 massive star in its final stage of evolution.  In addition, we assume that in the outer boundary the dissipated energy in the flow
 is totally advected:
\begin{eqnarray}
Q^+-Q^-\approx Q^+&\approx &\frac{\Sigma}{m_p} v_r
\left(\frac{de}{dr}-\frac{m_p p}{\rho^2} \frac{d\rho}{dr} \right) \nonumber \\
&\approx &T\frac{\Sigma}{m_p} v_r \frac{ds}{dr} \nonumber \\
&\approx &T\frac{\Sigma}{m_p}\xi v_r \frac{s}{r}, 
\end{eqnarray}
where $s$ is the specific entropy and $\xi$ is the parameter of the order of unity (see Chap. 8 of Kato et al. 1998).  The exact value of $\xi$ is not needed in this
 calculation.

With these boundary conditions, we solve the set of equations from the
outer boundary, using the fourth-order Runge-Kutta method.
\section{RESULTS}
In this section we present our results of the structure of hypercritical accretion flow with various mass accretion rates, and of the equilibrium curves.
  The flow structures with typical parameters are shown in Fig 1-3 for relatively low ($\dot{M}=0.01-0.08M_{\odot}~{\rm s}^{-1}$), moderately high
 ($\dot{M}=0.1-0.8M_{\odot}~{\rm s}^{-1}$), and very high ($\dot{M}=1.0-10.0M_{\odot}~{\rm s}^{-1}$) mass accretion rates, respectively.
  We can see from them that the profiles of density and temperature are highly dependent on mass accretion rates.  In the following, we see dominant
 physics in each of accretion regimes

\subsection{Cases of $\dot{M}=0.01-0.08M_{\odot}~{\rm s}^{-1}$}
First, we show the structures of the disks with $\dot{M}=0.01-0.08M_{\odot}~{\rm s}^{-1}$ in Fig 1. 
When disk temperature reaches $\sim {\rm MeV}$ around $r\sim 100r_g$ the photodissociation of nuclei occurs. We can see that the density profiles
 get slightly steeper, and the temperature profiles are flatter at $r<100r_g$ than in the other region (Fig 1).  These kinds of behavior can be
 understood as the effects of efficient
 cooling.  In the region of photodissociation, viscously dissipated energy is efficiently consumed rather than advected inward, and the
 flows are effectively cooled.  Therefore, the disks become thinner due to lower temperature and pressure, and densities get higher compared with
 the case in which dissociation is not taken into account.  The regions of photodissociation appear around $r\sim 100r_g$ with arbitrary mass
 accretion rates.  In the photodissociation region, the degeneracy parameter of electrons $\eta_e$ rises because of the photodissociation cooling. 

\subsection{Cases of $\dot{M}=0.1-0.8M_{\odot}~{\rm s}^{-1}$}
Next we show in Fig 2. the disk structures of the disk with moderately high mass accretion rate ($\dot{M}=0.1-0.8M_{\odot}~{\rm s}^{-1}$).  Around $r\sim 30r_g$,
  the density profiles get steeper and the temperature profiles get flatter, as seen in the photodissociation region.  Both of these are due to efficient
 neutrino cooling.  In this region, the gradient of
 entropy profile get positive (see lower-middle panel of Fig 2).  Almost the same argument as in the photodissociation region can be applied in this region.
  Namely, there should be effective cooling operating.

  In Fig 4, the ratios of neutrino cooling flux ($F_-$)to
 viscous heating rate per unit disk surface area ($F_+$)are shown.  We can see that, with moderately high mass accretion rate, there appears the region
 where $F_-/F_+$ gets drastically high or even exceeds unity (i.e. larger amount of energy than that viscously dissipated is emitted as neutrino per
 unit surface area).  Around that region, due to the sufficiently high density and temperature, thermal neutrino emissions become very efficient.
  Therefore, the accretion flow is well cooled there, and the profiles of density and temperature show the similar behavior as that in photodissociation
 region.  Note that we have always $Q_{\rm rad}/Q_{\rm vis}\sim 1$ in the standard disks.  In NDAF, excess cooling is compensated by advective heating.

  Moreover, electron fraction $Y_e$ dramatically decreases inward in this region.  This is because electron capture process
 ($e+p\rightarrow n+\nu_e$) becomes efficient with sufficiently high temperature ($\gtrsim 1{\rm MeV}$) and high electron degeneracy ($\eta _e\gtrsim 1$).
  Around there, $Y_e$ drops to as low as $\sim 0.1$, which means that the inner region of NDAF with moderate mass accretion rate is neutron-rich.

\subsection{Cases of $\dot{M}=1.0-10.0M_{\odot}~{\rm s}^{-1}$}
For very high mass accretion rate ($\gtrsim 3.0M_{\odot}~{\rm s}^{-1}$), we can see another interesting feature in the innermost
 region of an accretion flow, where the density profile becomes flatter, and the temperature profile becomes steeper than in the case with
 intermediate mass accretion rate (i.e. $\dot{M}\sim 0.3-1.0M_{\odot}~{\rm s}$).  This means that neutrino cooling which is efficient in lower
 accretion rate disk ceases to be powerful because the flow becomes thick for neutrinos.  The upper panel of Fig. 5 shows the profiles of the optical depth of
 $\nu_e$ in the disks with very high mass accretion rates.  According to this figure, in the innermost region, the accretion flows can be thick
 ($\tau_{\nu_e}\gtrsim 1$)with respect to neutrinos when mass accretion rates are larger than $\sim 3.0M_{\odot}~{\rm s}^{-1}$.  This makes an obvious
 change in the electron fraction profiles.  If neutrinos are trapped inside the flow, neutronization ($p+e^-\rightarrow n+\nu_e$) are effectively
 prevented because of the inverse reaction.  As a result, $Y_e$, which decreases inward in the efficient neutrino-cooled region, no longer decreases
 but rather increases inward in the
 neutrino-trapping region, and takes the value around $\sim 0.2$ in the vicinity of inner edge.  The steepness of density profiles
 or the flatness of temperature profiles are not so conspicuous as the cases with lower mass accretion rates.  The reason of this difference can be seen
 in the plots of free nucleon fraction (the lower-right panel of Fig 3).  For high mass accretion rates, the density gets high and so photodissociation
 of nuclei occurs rather gradually inward.  As a result, the peculiar features of density and temperature profiles are smeared out.

We can see other interesting features in the flow with even higher mass accretion rates.  In the innermost region, entropy per baryon and electron
 degeneracy parameter take approximately the same values regardless of mass accretion rates.  Such trends of physical quantities appear when
 the accretion flow gets thick with respect to neutrinos (see also Fig.5).  

Before ending this subsection, we should mention the oscillations
which appear conspicuously on the curves of temperature, degeneracy
parameter, and neutrino optical depth.  These features can be
explained in the following way.

In the outer region of the disk, most baryons are in the form of
nuclei, and the main opacity for the neutrinos is from the coherent
scattering by nuclei.  Around the radius where many oscillations on
the curves appear, the photodissociation of nuclei occurs and the
optical depth with respect to neutrinos drops inward suddenly.  As far
as the mass accretion rate is not so high, the optical depth is much
smaller than unity even if the plenty of heavy nuclei exist.  However,
is the mass accretion rate is sufficiently high, even in the outer
region the optical depth can be as large as the order of unity, and
due to the dissociation the accretion flow become suddenly
neutrino-thin inward.  Then the neutrino cooling become efficient and
the temperature can get lower inward, and the fraction of nuclei,
which is calculated from EoS table, gets suddenly increased inward.
This causes many oscillations on the curves of some physical
quantities.  Note that the physical quantities in the EoS table are
calculated assuming nuclear statistical equilibrium (NSE).  In our
case, this assumption is sometimes violated; if the timescale for
nucleons to attain NSE is longer than the accretion timescale the
composition of nuclei in the flow is expected to deviate from that
evaluated from the EoS table.  In order to calculate the nuclear
composition more precisely, it is necessary to solve the nuclear
reaction equations, which is beyond the scope of this work.

\subsection{Nuclear Composition of Hypercritical Accretion Flows}
Now that we solve the structures of hypercritical flows with arbitrary mass accretion rates, we can see how the composition in the disk depends on
 the radius and the mass accretion rate.  In the previous studies, only $\alpha$-particles and their dissociation were considered in the outer
 region of the disk, and the amount of $\alpha$-particles was estimated assuming nuclear statistic equilibrium.  This procedure, however,
 missed the existence of heavier nuclei (such as Fe, Ni, etc.), and so they could not take into account photodissociation
 cooling correctly, and could not follow the nuclear composition and the electron fraction consistently.  In our calculations, these points are
 greatly improved by employing the realistic EOS by Lattimer \& Swesty (1991)and properly considering the existence of nuclei other than He.

  Fig. 6 show the contour plots of electron fraction, $\alpha$-particle fraction, and heavy nucleon fraction on $(r,~\dot{M})$.  We can see that
 the existence of heavy nuclei is not negligible around $r\sim 100r_g$.  Heavy nuclei also affect the opacity of NDAFs with very high mass accretion rate.
   In the lower panel of Fig. 5, scattering optical depths with respect of electron neutrinos are shown with various mass accretion rates, and we can see that
 the scattering opacity by nuclei is not negligible in NDAFs with sufficiently high mass accretion rate.  This fact has not been mentioned in previous
 studies, in which the existence of heavy nuclei is neglected.  

\subsection{Stability Analysis of Hypercritical Accretion Flows}
We can draw the thermal equilibrium curves of NDAFs on $(\Sigma,T)$
  and $(\Sigma,\dot{M})$ plane by fixing some typical radii.
  These curves are helpful in checking qualitively the thermal and
  viscous stability of accretion flows (e.g. Di Matteo et al. 2002).  If there is an S-shaped sequence in the
 equilibrium curves on $(\Sigma,~T)/(\Sigma,~\dot{M})$, the disk is thermally/viscously unstable (Kato et al. 1998).  We expect some features in the
 equilibrium curves when dominant physics alternates others.  For
  example, S-shape appears when disks are cool enough for hydrogens to
  recombine in the case of dwarf-novae.
  Another S-shape appears when radiation pressure dominates over gas pressure.  We show those equilibrium curves
 in Fig 7 and 8 for $r=10r_g$, $30r_g,$ and $100r_g$.  These figures show that the NDAFs whose structures we calculate are thermally and viscously stable.
  This is because dissociation of nuclei does not induce drastic changes of physics dominating the disk structure.  However, we can find a remarkable features in the equilibrium curves.  At smaller radii,
 the equilibrium curves on the $(\Sigma,~T)$ plane have negative gradient part in a certain surface mass density range, and those on the
 $(\Sigma,~\dot{M})$ plane become comparatively flat with the same $\Sigma$.  This behavior corresponds to the efficient neutrino cooling
 in the disk with moderate mass accretion rate.  As this is the opposite tendency compared with the case of unstable disk, this feature implies that
 efficient neutrino cooling makes NDAF rather stable both thermally and viscously.  Such feature has not
 been found in the equilibrium curves of accretion flows with lower mass accretion rate ($\dot{M}\sim \dot{M}_{\rm Edd}$).  

\subsection{Neutrino Luminosity from Hypercritical Accretion Flows}
Fig. 9 shows the sum of neutrino luminosities of all flavors, electron neutrino luminosity and anti-electron neutrino luminosity (upper panel),
 and efficiency of neutrino emission ($L_{\nu}/\dot{M}c^2$, lower panel)as the function of mass accretion rate.  According to this plot, the energy
 conversion from the accretion flow to neutrino emissions is most efficient when mass accretion rate is $\sim 0.3M_{\odot}{\rm sec}^{-1}$.  Below
 this rate density and temperature in the accretion flow are not high enough to emit neutrinos efficiently, and above this rate the inner region of
 the disk gets thick with respect of neutrinos and so the number of neutrinos which escape the disk becomes smaller.

Note that, though we have taken into account the effects of heavy
nuclei in the analyses, the total neutrino luminosities do not differ
significantly from those calculated with the same mass accretion rate
in the previous studies, because most of neutrinos are emitted from
the innermost region of the disk ($r<10r_g$), while the
photodissociation of heavy nuclei occurs i the outer part of the disk
($r\sim 100r_g$). 
\section{DISCUSSIONS}
In this paper, we investigate the structure of hypercritical accretion flows as the central engine of GRBs with various mass accretion rates,
 and the thermal and viscous stability by drawing the thermal equilibrium curves on ($\Sigma,T$) and ($\Sigma,\dot{M}$) plane, respectively.
  We have found that the photodissociation of nuclei,
 efficient neutrino cooling, and neutrino trapping cause some remarkable changes to the disk structure.  In the calculation, we treat appropriately
 the composition of nuclei and the neutrino optical depths by adopting a realistic equation of state and various kinds of neutrino reaction.  
  By doing this, we firstly include the neutrino emissions and scattering by heavy nuclei, whose amount is not negligible around $r\sim 100r_g$
 assuming nuclear statistical equilibrium.  For relatively low mass accretion rates, remarkable differences from previous works do not appear in
 our calculations.  However, for very high mass accretion rate, even in the outer region ($r\sim 100r_g$)neutrino optical depths become of the
 order of unity.  This is because of the large opacity due to the coherent neutrino scattering from heavy nuclei (See the lower panel of Fig. 5).
  From these facts, we can conclude that the existence and dissociation of heavy nuclei remarkably affect the disk structure and neutrino emission,
 and this has not been mentioned in the previous works.

  We have also investigated the neutron-to-proton ratio in hypercritical accretion flows by taking into account the neutrino trapped inside the
 flows.  For moderately high mass accretion rates ($\dot{M}\sim 0.3-1.0M_{\odot}~{\rm s}^{-1}$), because of high density and efficient photodissociation
 cooling, the electron degeneracy parameter drastically increases inward around $r\sim 100r_g$.  In addition, the temperature becomes so high that
 many electrons have enough energy to be captured by protons.  Then in the inner region electron capture
 proceeds as the most efficient neutrino emission process, and then the electron fraction $Y_e$ decreases inward down to$\sim 0.1$.  This tendency
 is positive for the acceleration of an outflow that launches from around the innermost region of the disk; if the outflow picks up the disk
 material and becomes neutron-rich, the outflow will be accelerated more efficiently compared to the case that it picks up the same amount of disk material
 but neutron-poor, because neutrons do not interact with other particles electrostatically and will decouple while protons and electrons are accelerated
 (e.g. Vlahakis et al. 2003).
  Moreover, the disk outflow is considered to be a potential site for nucleosynthesis, and the conditions with large neutron-to-proton ratio is
 favorable to $r$-process (Surman et al. 2006).

On the other hand, if mass accretion rates are very high ($\dot{M}\gtrsim 3.0M_{\odot}~{\rm s}^{-1}$), in the innermost
 region $Y_e$ increases inward up to $\sim 0.2$.  This is because electron-neutrino optical depth becomes very high and electron neutrinos are trapped
 inside the accretion flow.  The same kind of features are seen in calculations of core-collapse supernova explosion, in which neutrino trapping
 due to the coherent neutrino scattering opacity by heavy nuclei stops the neutronization of stellar material because neutrinos become degenerated.
  Moreover, as seen in Fig. 8, the efficiency of neutrino emission from a whole disk surface peaks around $\dot{M}\sim 0.3M_{\odot}~{\rm s}^{-1}$
 and for larger mass accretion rates, $L_{\nu}/(\dot{M}c^2)$ is suppressed because of neutrino-thickness of the accretion flows.  These results imply
 that the accretion disks with too high mass accretion rate are not always favorable for the production of outflows with large Lorentz factor,
 which are considered to be necessary to explain observed GRB emission features.  
 
Our study also gives some implications to the possibility of energy deposition via neutrino pair annihilation ($\nu +\bar{\nu }\rightarrow e^- + e^+$)
 above the neutrino-cooled disk, which make a hot and baryon-poor fireball.  In order to estimate the energy deposition rate, we should calculate
 both the energy flux of $\nu_e$ and $\bar{\nu}_e$
 from the disk.  We investigate the effects of neutron-to-proton ratio, electron degeneracy, and neutrino trapping on the flavor of neutrinos mainly
 emitted from the disk surface.  As seen in Fig. 8, the luminosity of neutrinos from the whole disk surface is around
 $L_{\nu}\sim 10^{52}{\rm erg}~{\rm s}^{-1}$, and there is little difference between the luminosities of $\nu_e$ and $\bar{\nu}_e$, which had often been
 assumed in the previous studies on neutrino pair annihilations above the neutrino-dominated accretion disk (e.g. Gu et al. 2006).
  So we can say that their estimates for energy deposition rate above the disk are essentially unchanged.  For the precise calculation, however,
 we should know the energy spectrum of emitted neutrinos because the cross section of pair annihilation depends on the energy of neutrinos and antineutrino.
  We are now investigating the possibility of neutrino spectra from neutrino-thick accretion disks to deform from Fermi-Dirac distribution and
 its influence on the energy deposition rates (Kawabata et al. in preparation).

Finally, we mention the stability of hypercritical accretion flows.  In this paper we have obtained the thermal equilibrium curves of the accretion flow
 at several radii where the dominant physics operating the flow change, and have found that hypercritical accretion flows are both thermal and viscously
 stable, even at the radius where the photodissociation of heavy nuclei occurs and the nuclear composition and opacity sources drastically change.
  Therefore, we cannot explain the time-variable behavior in the context of these kinds of instabilities.
  We have not considered here the effect of gravitational instability of the disk (Toomre 1964).  As already indicated by many authors, the accretion
 flows with $\dot{M}\gtrsim 1M_{\odot}~{\rm s}^{-1}$ are gravitationally unstable in their outer part (e.g. Di Matteo et al. 2002).  Such disks can
 fragment into gravitationally bound blobs, and this can be responsible for X-ray flares observed in the early phase of GRB afterglows (Perna et al. 2006).
  In order to investigate the behavior of fragmented hypercritical accretion flows, one should reconstruct the time-dependent disk model taking into
 account the self-gravity, which is left for future work.

We are grateful to H. Suzuki for providing us the table of EoS by Lattimer \& Swesty.  We also thank S. Nagataki, S. Uematsu, and R. Takahashi for
 useful discussions and helpful comments.  The numerical calculations were carried out on Altix3700 BX2 at Yukawa Institute for Theoretical Physics
 in Kyoto University.  NK is supported in part by Research Fellowships of the Japan Society for the Promotion of Science for Young Scientists.

\appendix
\section{Computations of Neutrino Emission Rates and Opacities}
In this appendix we give the way to calculate the neutrino emission rates and neutrino optical depth in an NDAF along the vertical direction.
  Most of the expressions we show below are covered by Bruenn (1985), Ruffert et al. (1996) and Burrows \& Thompson (2002).

The number emission rate and energy emission rate for electron capture by protons ($p+e^- \rightarrow n+\nu_e$)and those for positron capture by
 neutrons ($n+e^+ \rightarrow p+\bar{\nu_e}$)are given by
\begin{eqnarray}
R_{p+e^- \rightarrow n+\nu_e}&=&\frac{G_F^2}{2\pi^3 \hbar^3 c^2}(1+3g_A^2) \eta_{pn} \int_Q^{\infty}dE_e E_e\sqrt{E_e^2-m_e^2 c^4} \nonumber \\
&&\times (E_e-Q)^2 \frac{1}{e^{(E_e-\mu_e)/k_B T}+1}, \\
q^-_{p+e^- \rightarrow n+\nu_e}&=&\frac{G_F^2}{2\pi^3 \hbar^3 c^2}(1+3g_A^2) \eta_{pn} \int_Q^{\infty}dE_e E_e\sqrt{E_e^2-m_e^2 c^4} \nonumber \\
&&\times (E_e-Q)^3 \frac{1}{e^{(E_e-\mu_e)/k_B T}+1}, \\
R_{n+e^+ \rightarrow p+\bar{\nu_e}}&=&\frac{G_F^2}{2\pi^3 \hbar^3 c^2}(1+3g_A^2) \eta_{np} \int_{m_e c^2}^{\infty}dE_e E_e\sqrt{E_e^2-m_e^2 c^4} \nonumber \\
&&\times (E_e+Q)^2 \frac{1}{e^{(E_e+\mu_e)/k_B T}+1}, \\
q^-_{n+e^+ \rightarrow p+\bar{\nu_e}}&=&\frac{G_F^2}{2\pi^3 \hbar^3 c^2}(1+3g_A^2) \eta_{np} \int_{m_e c^2}^{\infty}dE_e E_e\sqrt{E_e^2-m_e^2 c^4} \nonumber \\
&&\times (E_e+Q)^3 \frac{1}{e^{(E_e+\mu_e)/k_B T}+1},
\end{eqnarray}
respectively, where $G_F=2.302\times 10^{-22} {\rm cm}~{\rm MeV}^{-1}$ is the Fermi constant, $g_A\simeq 1.23$ is the axial
 vector coupling constant, $Q\simeq 1.29{\rm MeV}$ is the difference of rest mass energy between a neutron and a proton,
 $m_e$ is the electron mass, and $\mu_e$ is the chemical potential of electrons. Here, $\eta_{pn}$ and $\eta_{np}$ indicate the inclusion of
 Fermion blocking effects in the nucleon phase spaces and are defined as
\begin{eqnarray}
\eta_{pn}=\frac{n_n-n_p}{{\rm exp}(\eta_n -\eta_p)-1}, \\
\eta_{np}=\frac{n_p-n_n}{{\rm exp}(\eta_p -\eta_n)-1},
\end{eqnarray}
respectively.  Here, $n_n$ and $n_p$ are the number density of neutrons and protons, respectively, and $\eta_n$ and $\eta_p$ are the degeneracy
 parameters neutrons and protons, respectively.  These quantities, $\eta_{pn}$ and $\eta_{np}$, are equal to $n_p$ and $n_n$, respectively,
 in the nondegenerate regime (where blocking is not important), but when the nucleons are degenerate, those values become smaller than
 $n_n$ and $n_p$, respectively. 

The number emission rate and energy emission rate for electron capture by nuclei ($A + e^- \rightarrow A^{\prime }+\nu_e$) depend on the
 internal proton number $Z$ and the internal neutron number $N$, and are given by
\begin{eqnarray}
R_{A+e^- \rightarrow A^{\prime}+\nu_e}&=&\frac{G_F^2}{2\pi^3 \hbar^3 c^2}g_A^2 \frac{2}{7} N_p(Z) N_h(N) n_A \nonumber \\
&&\times \int_{Q^{\prime}}^{\infty}dE_e E_e\sqrt{E_e^2-m_e^2 c^4}(E_e-Q^{\prime})^2 \frac{1}{e^{(E_e-\mu_e)/k_B T}+1}, \\
q^{-}_{A+e^- \rightarrow A^{\prime}+\nu_e}&=&\frac{G_F^2}{2\pi^3 \hbar^3 c^2}g_A^2 \frac{2}{7} N_p(Z) N_h(N) n_A \nonumber \\
&&\times \int_{Q^{\prime}}^{\infty}dE_e E_e\sqrt{E_e^2-m_e^2 c^4}(E_e-Q^{\prime})^3 \frac{1}{e^{(E_e-\mu_e)/k_B T}+1},
\end{eqnarray}
where
\begin{eqnarray}
Q^{\prime}&\approx &\mu_n-\mu_p+\Delta , \\
N_p(Z)&=&\begin{cases}
0,&Z<20 \\
Z-20,&20<Z<28 \\
8,&Z>28,
\end{cases}
\\
N_h(N)&=&\begin{cases}
6,&N<34 \\
40-N,&34<N<40 \\
0,&N>40.
\end{cases}
\end{eqnarray}
  The quantity $\Delta\approx 3{\rm MeV}$ is the energy of the neutron $1f_{5/2}$ state above the ground state.

The number emission rates and the energy emission rates of $\nu_e$ or $\bar{\nu_e}$ by pair annihilation
 ($e^- + e^+ \rightarrow \nu_i +\bar{\nu_i}$) are given by
\begin{eqnarray}
R_{e^- + e^+ \rightarrow \nu_e +\bar{\nu_e}}&=&\frac{1}{2}\frac{(C_1+C_2)_{\nu_e \bar{\nu_e}}}{36}\frac{\sigma_0 c}{(m_e c^2)^2}
 \epsilon _{e^-} \epsilon_{e^+}, \\
q^-_{e^- + e^+ \rightarrow \nu_e +\bar{\nu_e}}&=&\frac{1}{2}\frac{(C_1+C_2)_{\nu_e \bar{\nu_e}}}{36}\frac{\sigma_0 c}{(m_e c^2)^2}
 (\tilde {\epsilon}_{e^-}\epsilon_{e^+}+\epsilon_{e^-}\tilde{\epsilon}_{e^+}), \\
R_{e^- + e^+ \rightarrow \nu_x +\bar{\nu_x}}&=&\frac{1}{2}\frac{(C_1+C_2)_{\nu_x \bar{\nu_x}}}{9}\frac{\sigma_0 c}{(m_e c^2)^2}
 \epsilon_{e^-}\epsilon_{e^+}, \\
q^-_{e^- + e^+ \rightarrow \nu_x +\bar{\nu_x}}&=&\frac{1}{2}\frac{(C_1+C_2)_{\nu_x \bar{\nu_x}}}{9}\frac{\sigma_0 c}{(m_e c^2)^2}
 (\tilde {\epsilon}_{e^-}\epsilon_{e^+}+\epsilon_{e^-}\tilde{\epsilon}_{e^+}),
\end{eqnarray}
where the weak interaction constants that appear in the above expressions are
\begin{eqnarray}
\sigma_0&=&1.76\times 10^{-44} {\rm cm}^{-3}, \\
(C_1+C_2)_{\nu_e \bar{\nu_e}}&=&(C_V-C_A)^2 + (C_V+C_A)^2, \\
(C_1+C_2)_{\nu_x \bar{\nu_x}}&=&(C_V-C_A)^2 + (C_V+C_A-2)^2,
\end{eqnarray}
with $C_V=1/2+2{\rm sin}^2 \theta _W$, ${\rm sin}^2 \theta_W\approx 0.23$, and $C_A=1/2$.

The emission rates of creation of $\nu_e$ and $\bar{\nu_e}$ by plasmon decay ($\tilde{\gamma} \rightarrow \nu_i + \bar{\nu_i}$) are given by
\begin{eqnarray}
R_{\tilde{\gamma} \rightarrow \nu_e + \bar{\nu_e}}&=&\frac{\pi^3}{6\alpha^{\ast }}C_V^2\frac{\sigma_0 c}{(m_e c^2)^2}\frac{(k_B T)^8}{(2\pi\hbar c)^6}
\gamma^6 e^{-\gamma} (1+\gamma), \\
q^-_{\tilde{\gamma} \rightarrow \nu_e +\bar{\nu_e}}&=&\frac{\pi^3}{6\alpha^{\ast}}C_V^2 \frac{\sigma_0 c}{(m_e c^2)^2}\frac{(k_B T)^9}{(2\pi\hbar c)^6}
\gamma^6 e^{-\gamma} (\gamma^2 +2\gamma +2), \\
R_{\tilde{\gamma} \rightarrow \nu_x + \bar{\nu_x}}&=&\frac{4\pi^3}{6\alpha^{\ast }}(C_V-1)^2\frac{\sigma_0 c}{(m_e c^2)^2}\frac{(k_B T)^8}{(2\pi\hbar c)^6}
\gamma^6 e^{-\gamma} (1+\gamma), \\
q^-_{\tilde{\gamma} \rightarrow \nu_e +\bar{\nu_e}}&=&\frac{4\pi^3}{6\alpha^{\ast}}(C_V-1)^2 \frac{\sigma_0 c}{(m_e c^2)^2}\frac{(k_B T)^9}{(2\pi\hbar c)^6}
\gamma^6 e^{-\gamma} (\gamma^2 +2\gamma +2),
\end{eqnarray}
where $\alpha^{\ast}\approx 1/137$ is the fine structure constant, and $\gamma\approx \gamma_0 \sqrt{(\pi^2+3\eta_e^2)/3}$ with $\gamma_0$ being related
 to the plasma frequency by $\gamma_0=\hbar \Omega_0/m_e c^2=5.565\times 10^{-2}$.

The absorption opacities of neutrinos can be derived from the above expressions of emissivity.  Using Kirchhoff's law, the spectral-integrated
 neutrino optical depths are given by
\begin{eqnarray}
\tau_{n,a,\nu_e}&=&\frac{R_{p+e^- \rightarrow n+\nu_e}+R_{e^- +A \rightarrow \nu_e + A^{\prime}}+R_{e^- +e^+ \rightarrow \nu_e +\bar{\nu_e}}
+R_{\tilde{\gamma}\rightarrow \nu_e +\bar{\nu_e}}}{c n_{\nu_e,0}}, \\
\tau_{E,a,\nu_e}&=&\frac{q^-_{p+e^- \rightarrow n+\nu_e}+q^-_{e^- +A \rightarrow \nu_e + A^{\prime}}+q^-_{e^-+e^+ \rightarrow \nu_e +\bar{\nu_e}}
+q^-_{\tilde{\gamma}\rightarrow \nu_e+\bar{\nu_e}}}{c u_{\nu_0,e}}, \\
\tau_{n,a,\bar{\nu_e}}&=&\frac{R_{n+e^+ \rightarrow p+\bar{\nu_e}}+R_{n \rightarrow p+e^-+\bar{\nu_e}}+R_{e^-+e^+\nu_e+\bar{\nu_e}}
+R_{\tilde{\gamma}\rightarrow \nu_e+\bar{\nu_e}}}{c n_{\bar{\nu_e},0}}, \\
\tau_{E,a,\bar{\nu_e}}&=&\frac{q^-_{n+e^+ \rightarrow p+\bar{\nu_e}}+q^-_{n \rightarrow p+e^-+\bar{\nu_e}}
+q^-_{e^- +e^+\rightarrow \nu_e+\bar{\nu_e}}+q^-_{\tilde{\gamma}\rightarrow \nu_e+\bar{\nu_e}}}{c u_{\bar{\nu_e},0}}, \\
\tau_{n,a,\nu_x}&=&\frac{R_{e^-+e^+ \rightarrow \nu_e+\bar{\nu_e}}+R_{\tilde{\gamma}\rightarrow \nu_e +\bar{\nu_e}}}{c n_{\nu_x,0}}, \\
\tau_{E,a,\nu_x}&=&\frac{q^-_{e^-+e^+ \rightarrow \nu_e+\bar{\nu_e}}+q^-_{\tilde{\gamma}\rightarrow \nu_e +\bar{\nu_e}}}{c u_{\nu_x,0}},
\end{eqnarray}
where $n_{\nu_i,0}$ and $u_{\nu_i,0}$ are the neutrino number density and energy density, respectively, that are spectral-integrated assuming
 Fermi-Dirac distribution.  

In calculating the scattering opacities of neutrinos, we consider nucleon scattering, electron scattering and heavy nucleus scattering, using
 Fermi-Dirac distribution function in spectral-averaging.  First, the nucleon scattering opacities are computed as
\begin{eqnarray}
\kappa_{j,s,\nu_i N}=C_{s,N}\sigma_0 \frac{\rho}{m_N} Y_{NN} \left( \frac{k_B T}{m_e c^2} \right)^2 \cdot \frac{F_{4+j}(\eta_{nu_i})}{F_{2+j}(\eta_{\nu_i})}.
\end{eqnarray}
where $C_{s,p}=[4(C_V-1)^2+5\alpha^2]/24$ for a proton and $C_{s,n}=(1+5\alpha^2)/24$ for a neutron with $\alpha\approx 1.25$.  For $j=0$ one
 obtains the opacities for neutrino number transport, and for $j=1$ the opacities for neutrino energy transport.  The variable $Y_{NN}$ means
 the nucleon number fraction in which Fermi blocking effects are taken into account, and here we define as
\begin{eqnarray}
Y_{NN}=\frac{Y_N}{1+(2/3){\rm max}(\eta_N,0)},
\end{eqnarray}
where $Y_N$ and $\eta_N$ are the number fraction and degeneracy parameter of each nucleon, respectively.

The electron scattering opacities are given by
\begin{eqnarray}
\kappa_{j,s,\nu_i e}\approx n_e\cdot \frac{3\sigma_0}{8}\left(\frac{T}{m_e c^2}\right)^2 \left( 1+\frac{\eta_e}{4} \right)
 \left[ (C_V+C_A)^2+\frac{1}{3}(C_V-C_A)^2 \right]\frac{F_{3+j}(\eta_{\nu_i})}{F_{2+j}(\eta_{\nu_i})},
\end{eqnarray}
where $n_e$ is the number density of electrons, and the weak interaction constants are $C_V=1/2+2{\rm sin}^2\theta_W$ for electron types,
 $C_V=-1/2+2{\rm sin}^2 \theta_W$ for $\mu$ and $\tau$ types, $C_A=1/2$ for
$\nu_e$, $\bar{\nu_{\mu}}$ and $\bar{\nu_{\tau}}$, and $C_A=-1/2$ for $\bar{\nu_e}$, $\nu_{\mu}$ and $\nu_{\tau}$.

The effects of neutrino-nucleus scattering will be important in the outer region of the disk, where heavy nuclei are not fully dissociated into
 nucleons.  The opacities are given by
\begin{eqnarray}
\kappa_{j,s,\nu_i A}\approx n_A\cdot\frac{\sigma_0}{16}\left(\frac{k_B T}{m_e c^2} \right)^2 \frac{F_{4+j}(\eta_{\nu_i})}{F_{2+j}(\eta_{\nu_i})},
\end{eqnarray}
where $n_A$ is the number density of nuclei with the atomic weight $A$. 

Finally, we mention about the choice of the degeneracy parameters of neutrinos.  In calculating neutrino emissivities and opacities according to
 the formulae shown above, one should choose the appropriate values of neutrino degeneracy parameters, $\eta_{\nu_i}$.
  We choose
\begin{eqnarray}
\eta_{\nu_x}&=&0, \\
\eta_{\nu_e}&=&\eta^{\rm ceq}_{\nu_e}\cdot \left[ 1-{\rm exp}(-\tau_{n,\nu_e}) \right], \\
\eta_{\bar{\nu_e}}&=&\eta^{\rm ceq}_{\bar{\nu_e}}\cdot \left[ 1-{\rm exp}(-\tau_{n,\bar{\nu_e}}) \right], 
\end{eqnarray}
with
\begin{eqnarray}
\eta^{\rm ceq}_{\nu_e}&=&-\eta^{\rm ceq}_{\bar{\nu_e}}=\eta_e+\eta_p-\eta_n.
\end{eqnarray}
With this choice, one can reproduce the neutrino number density and energy density at chemical equilibrium in the neutrino-thick limit. 

\clearpage
\vfill
\begin{figure}
  \begin{center}
    \begin{tabular}{ccc}
      \resizebox{50mm}{!}{\includegraphics{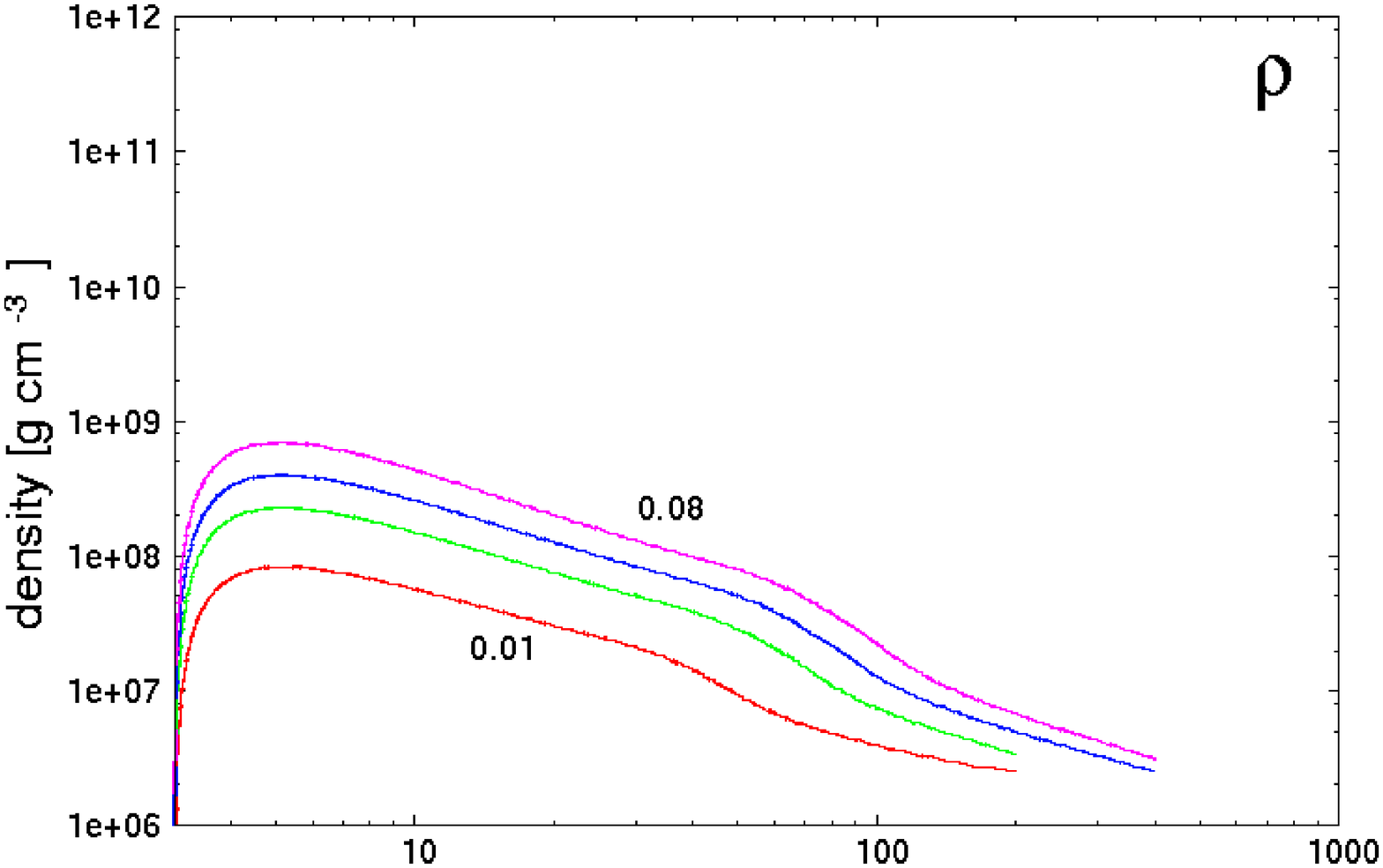}} &
      \resizebox{50mm}{!}{\includegraphics{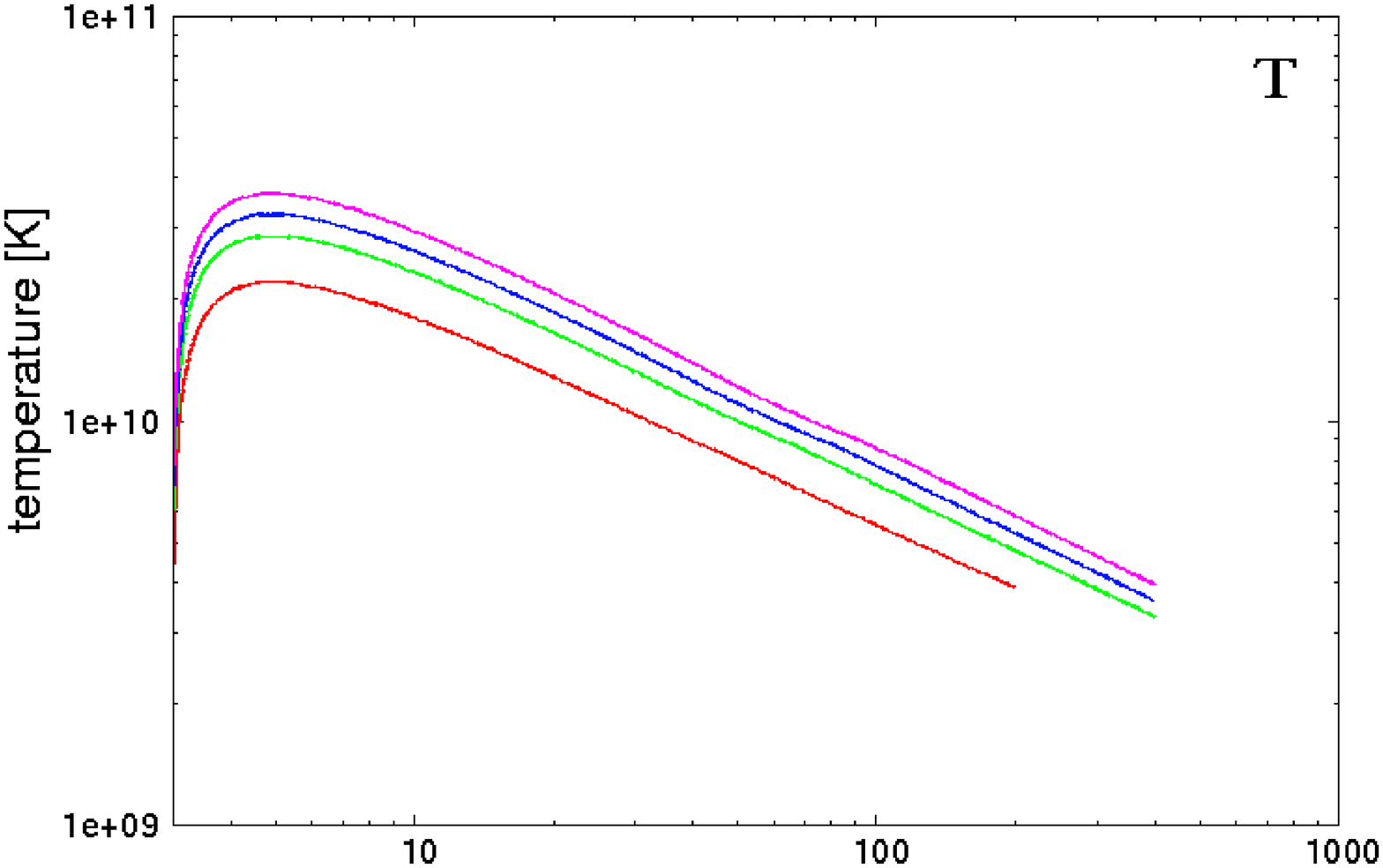}} &
      \resizebox{50mm}{!}{\includegraphics{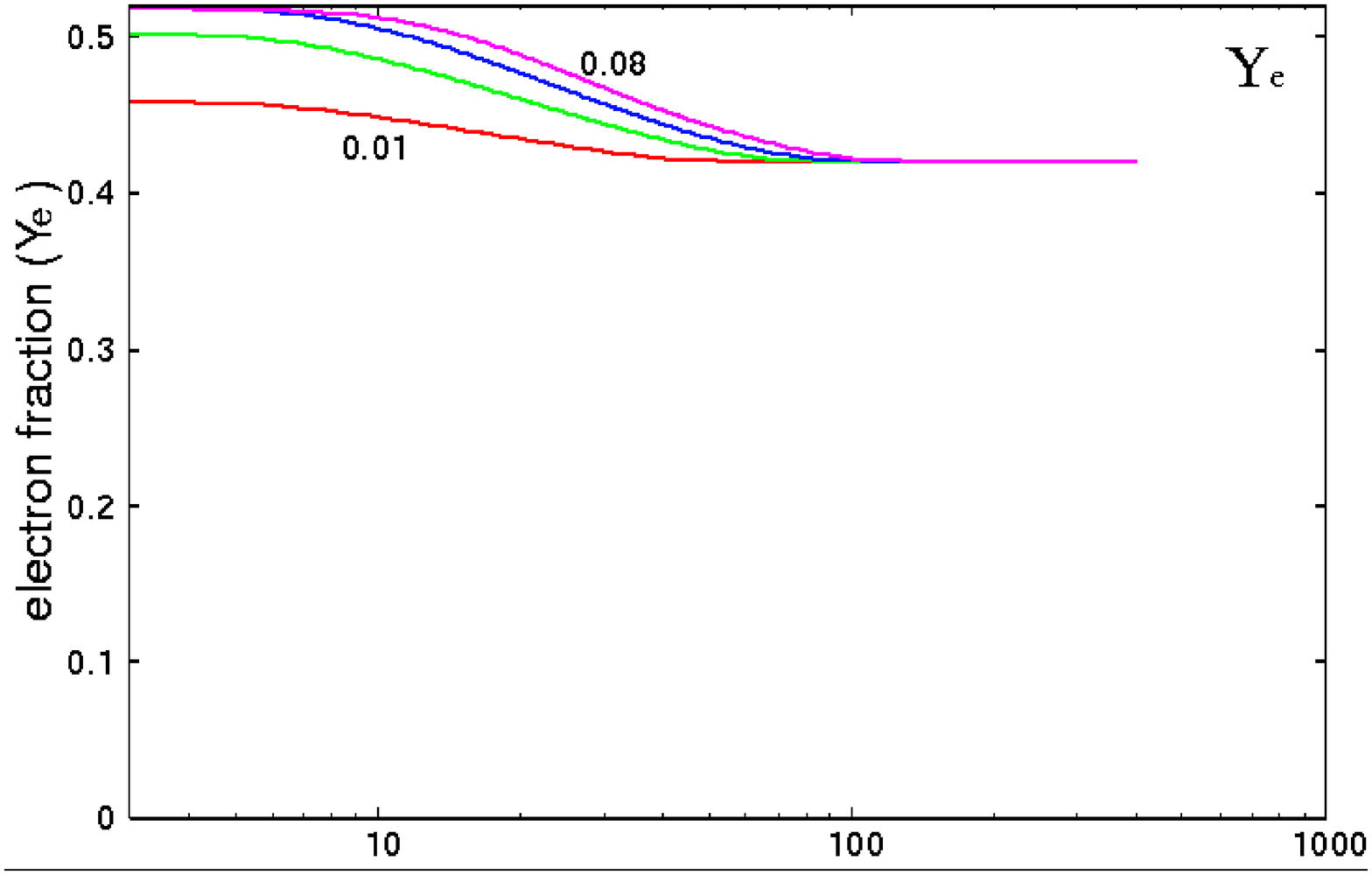}} \\
      \resizebox{50mm}{!}{\includegraphics{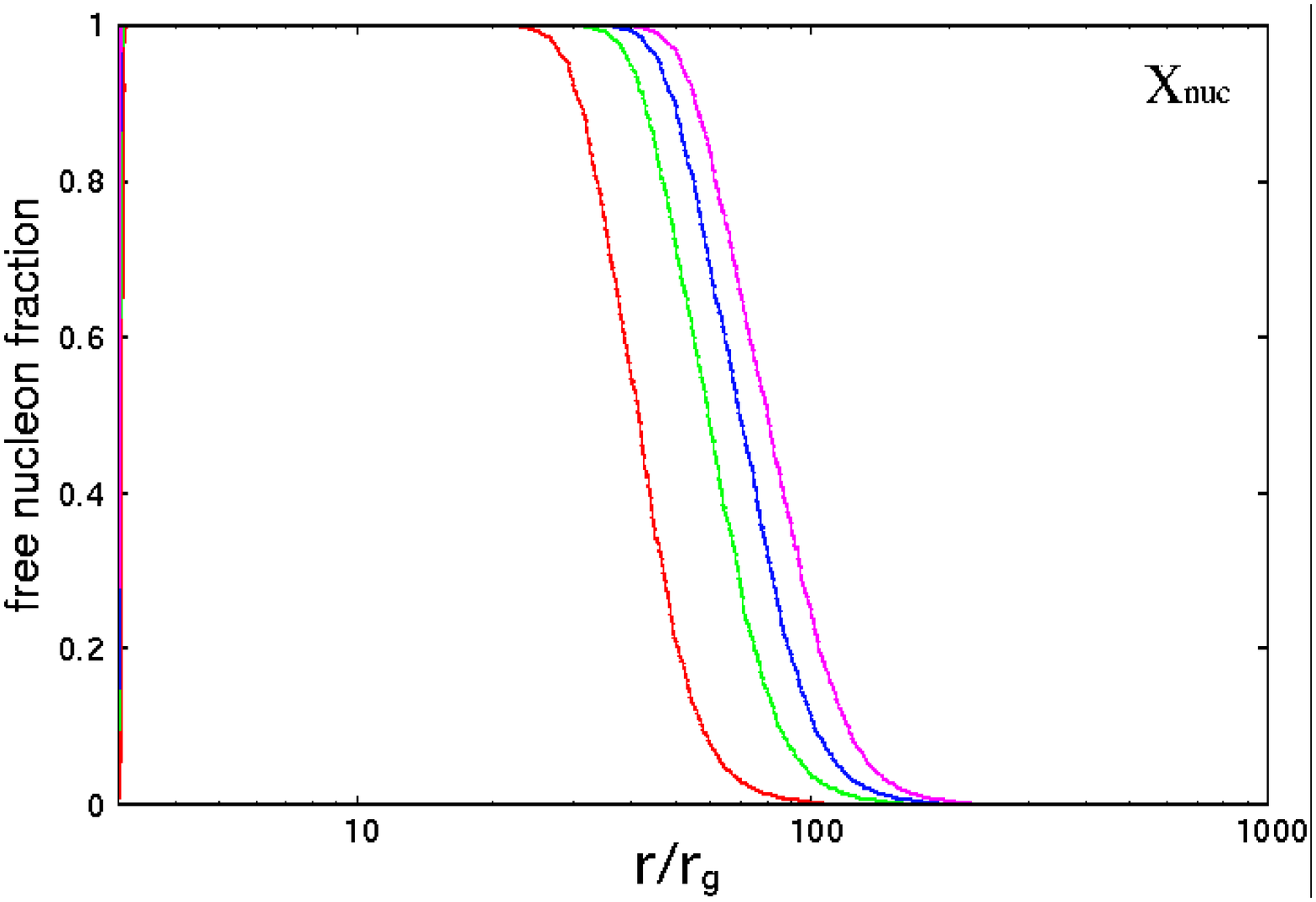}} &
      \resizebox{50mm}{!}{\includegraphics{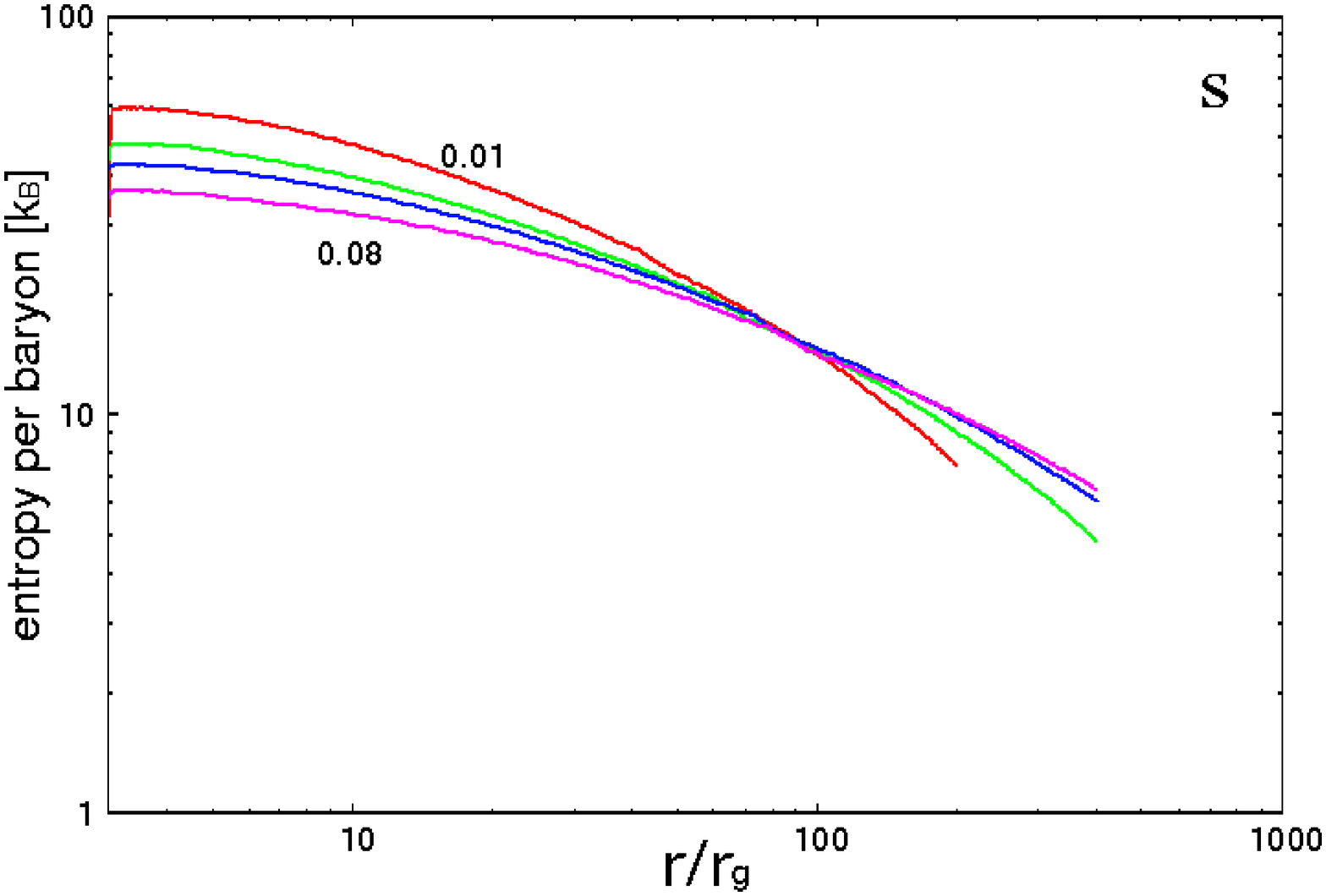}} &
      \resizebox{50mm}{!}{\includegraphics{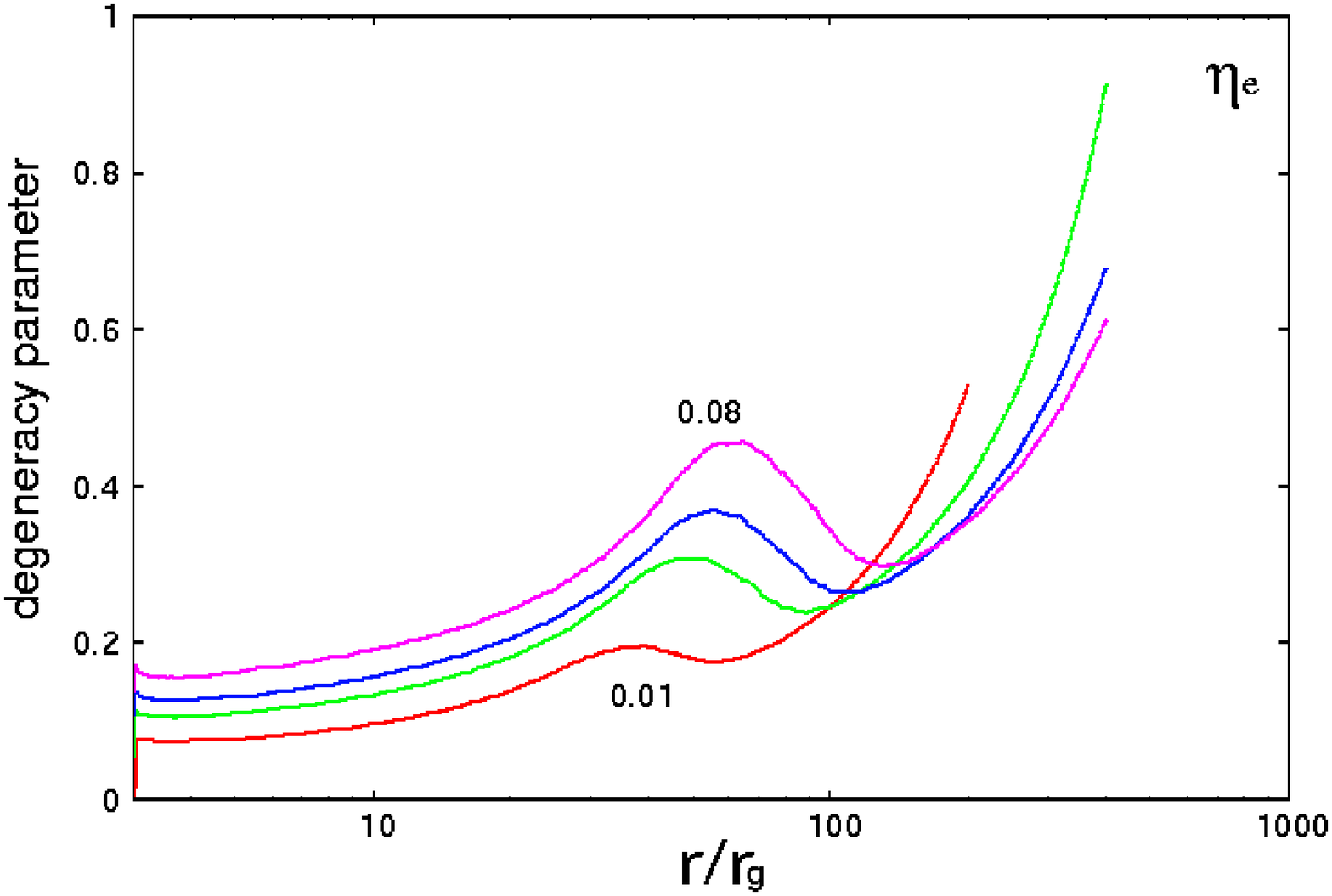}} \\
    \end{tabular}
    \caption{Profiles of density (upper-left), temperature (upper-middle), electron fraction (upper-right), free nucleon fraction (lower-left),
 entropy per baryon (lower-middle), and electron degeneracy parameter (lower-right)
 for the accretion flow with $\dot{M}=0.01$ ({\it red}), $0.03$ ({\it green}), $0.05$ ({\it blue}), and $0.08M_{\odot}~{\rm sec}^{-1}$ ({\it purple}).}
    \label{lowacc}
  \end{center}
\end{figure}
\vfill
\clearpage
\vfill
\begin{figure}
  \begin{center}
    \begin{tabular}{ccc}
      \resizebox{50mm}{!}{\includegraphics{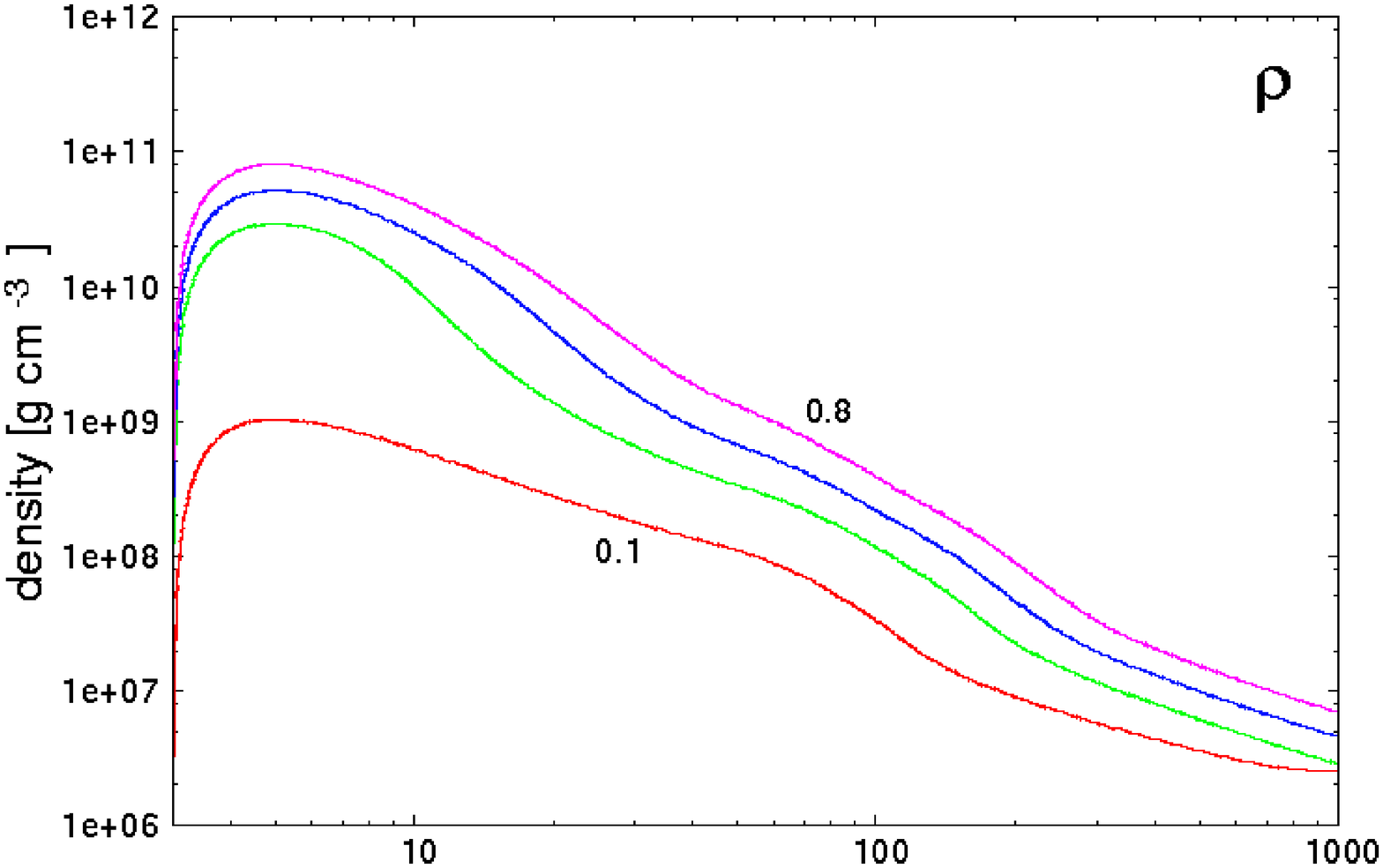}} &
      \resizebox{50mm}{!}{\includegraphics{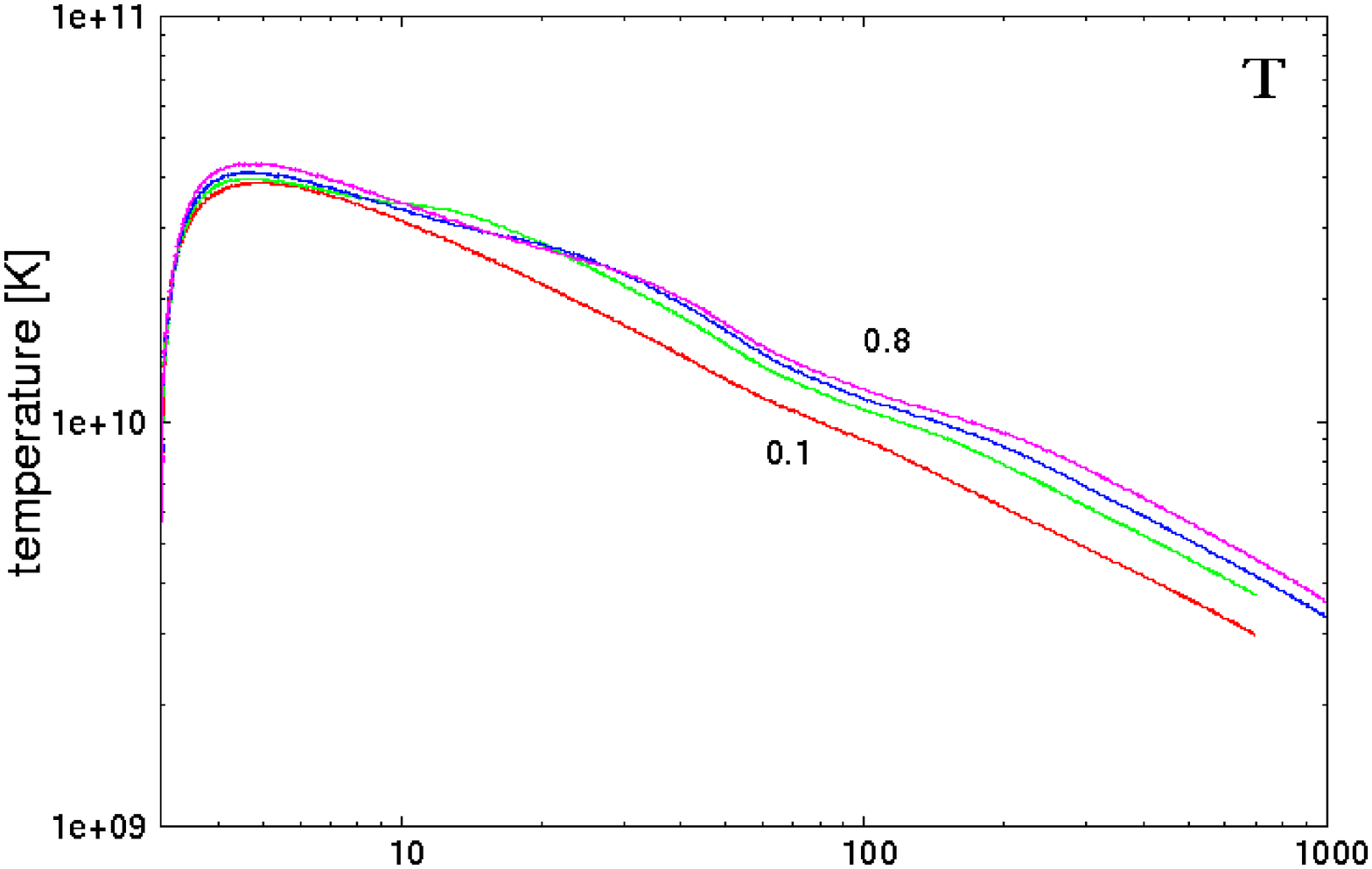}} &
      \resizebox{50mm}{!}{\includegraphics{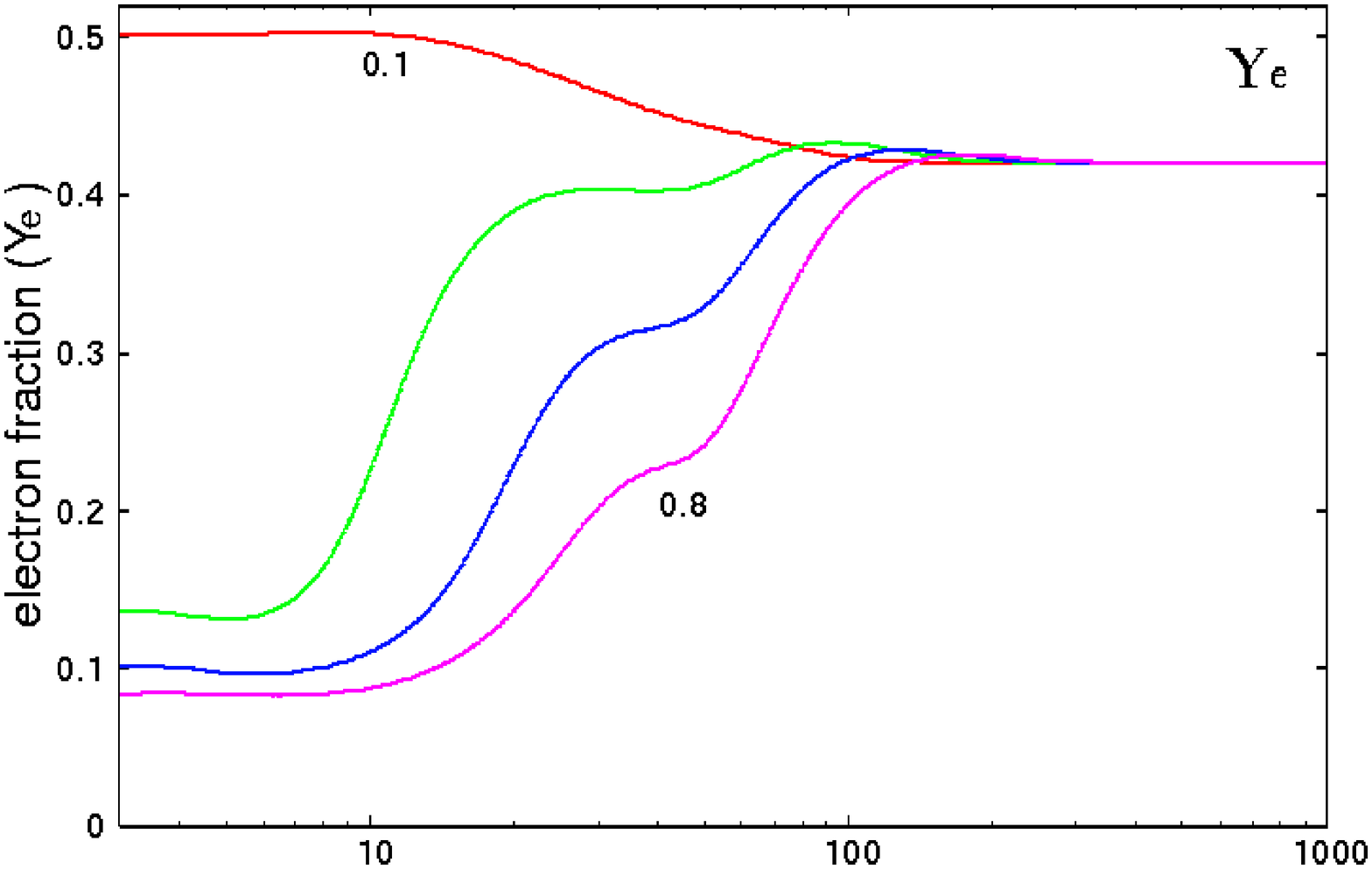}} \\
      \resizebox{50mm}{!}{\includegraphics{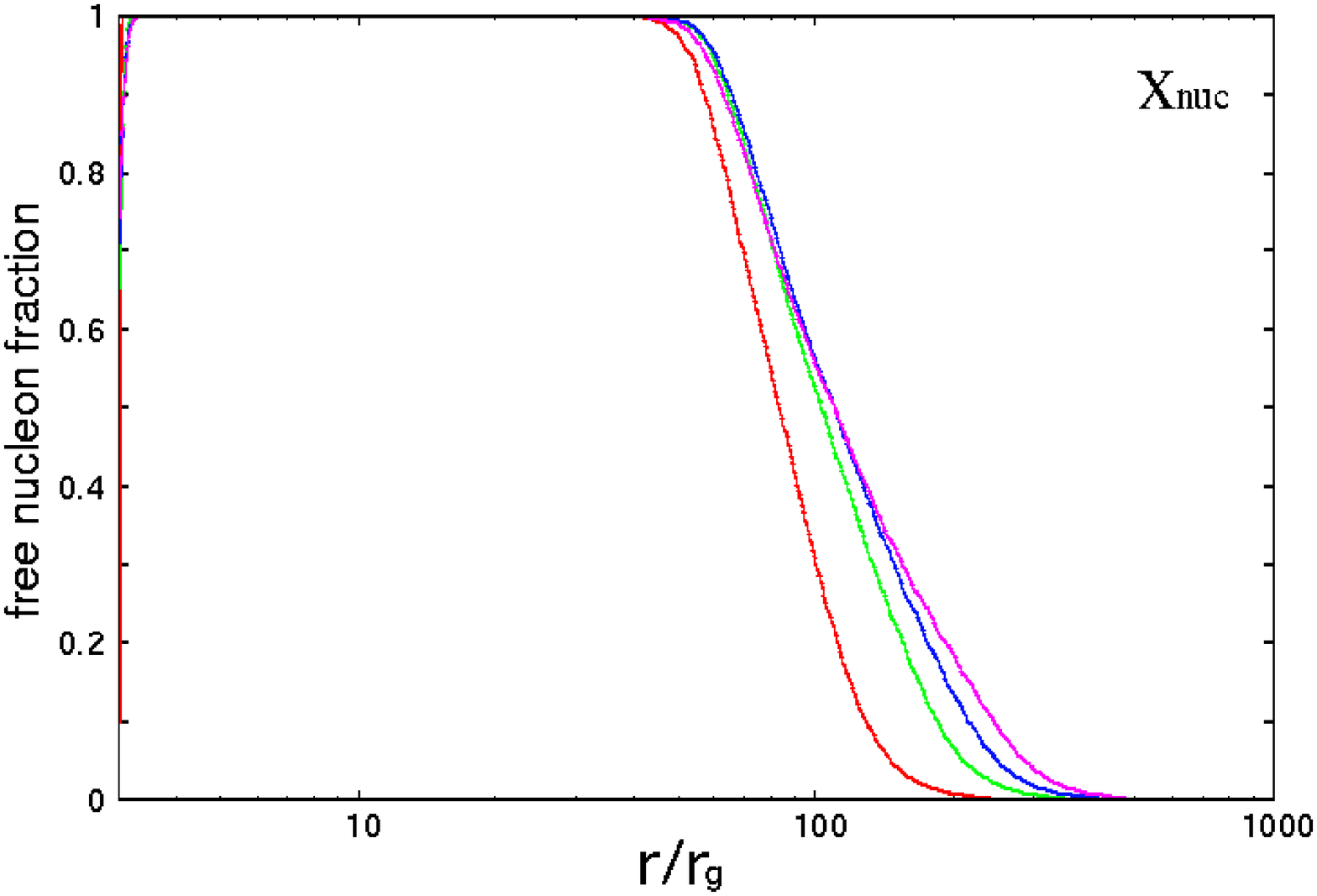}} &
      \resizebox{50mm}{!}{\includegraphics{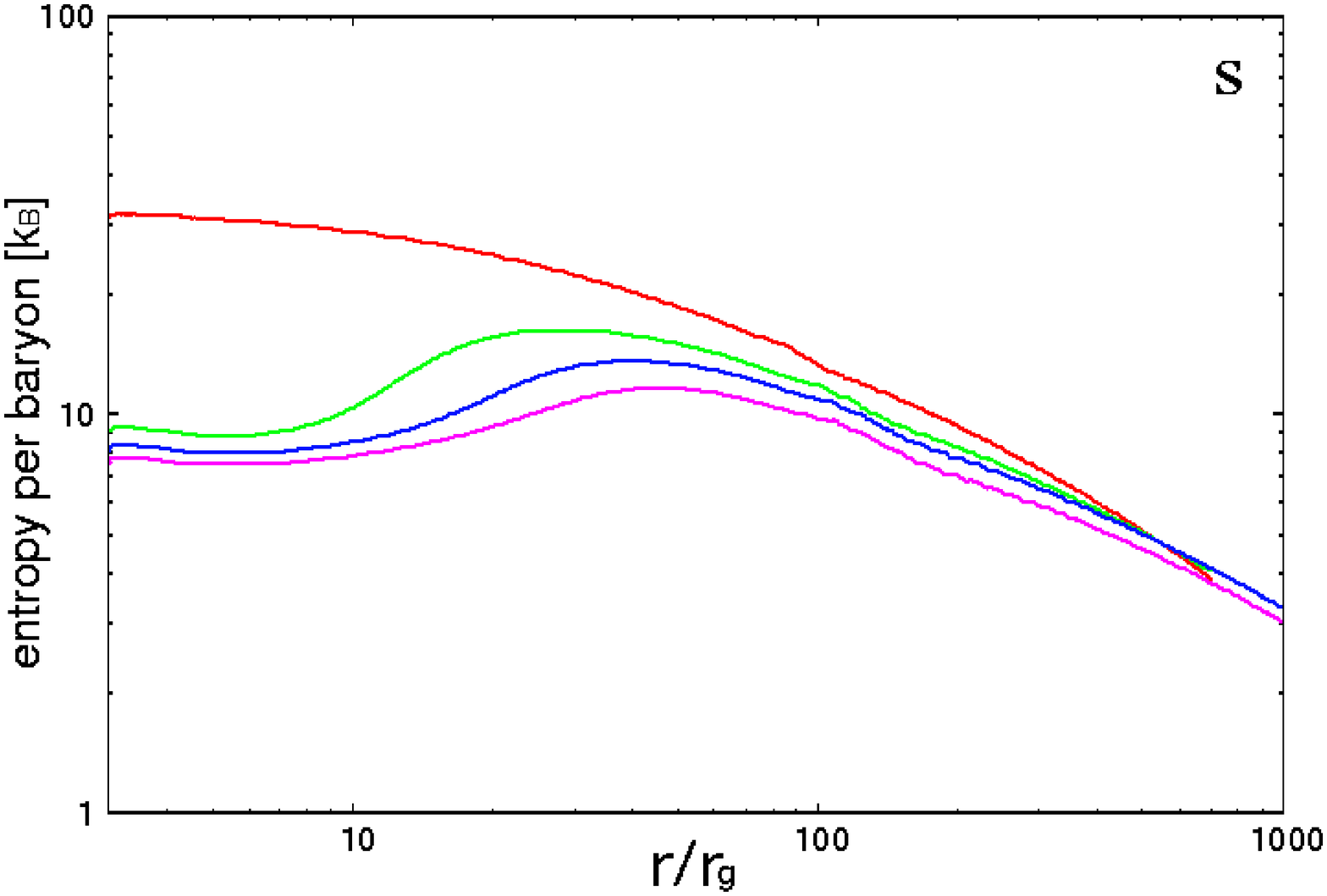}} &
      \resizebox{50mm}{!}{\includegraphics{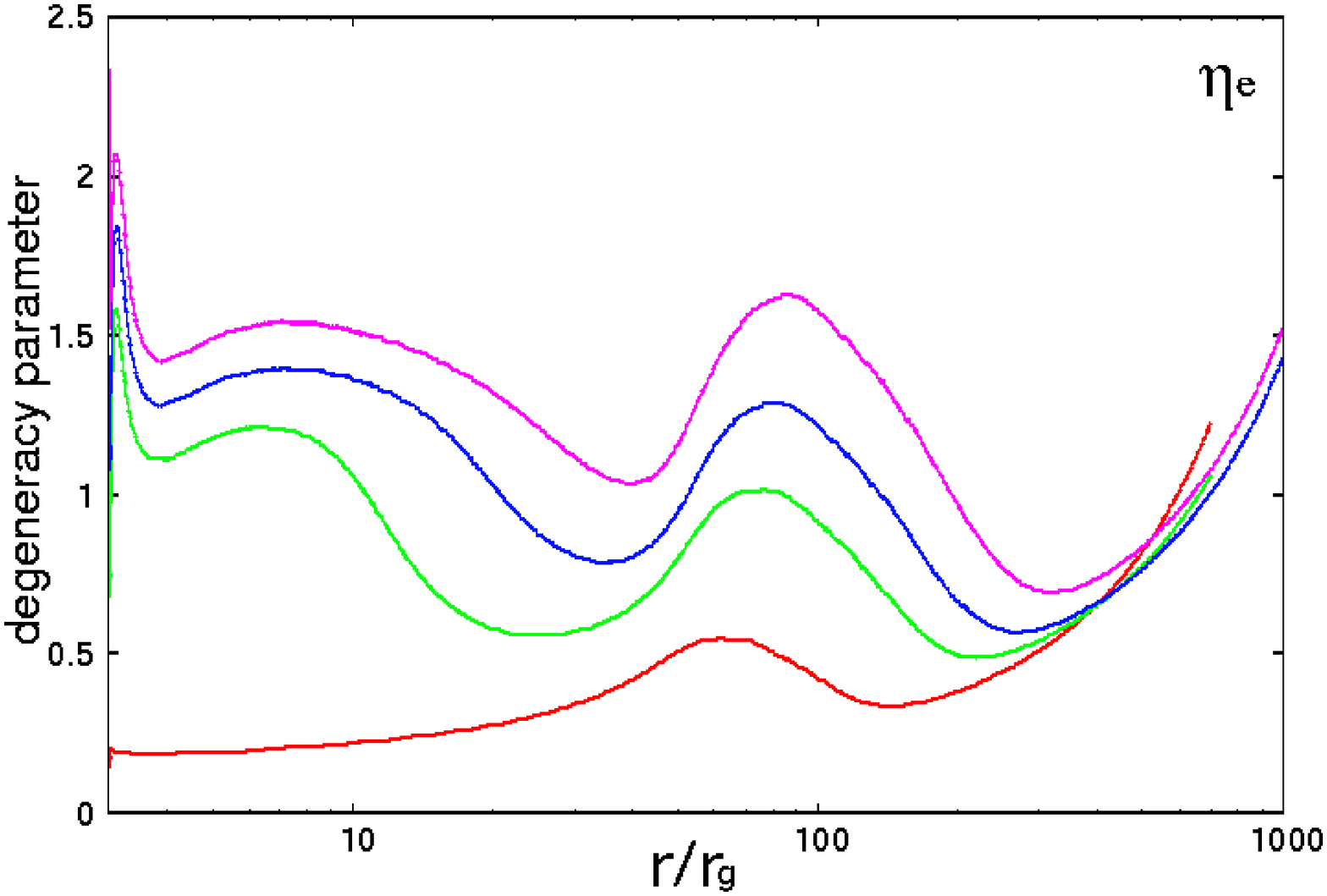}} \\
    \end{tabular}
    \caption{Same as Fig 1., but for the accretion flow with $\dot{M}=0.1$ ({\it red}), $0.3$ ({\it green}), $0.5$ ({\it blue}), and $0.8M_{\odot}~{\rm sec}^{-1}$ ({\it purple}).}
    \label{medacc}
  \end{center}
\end{figure}
\vfill
\clearpage
\vfill
\begin{figure}
  \begin{center}
    \begin{tabular}{ccc}
      \resizebox{50mm}{!}{\includegraphics{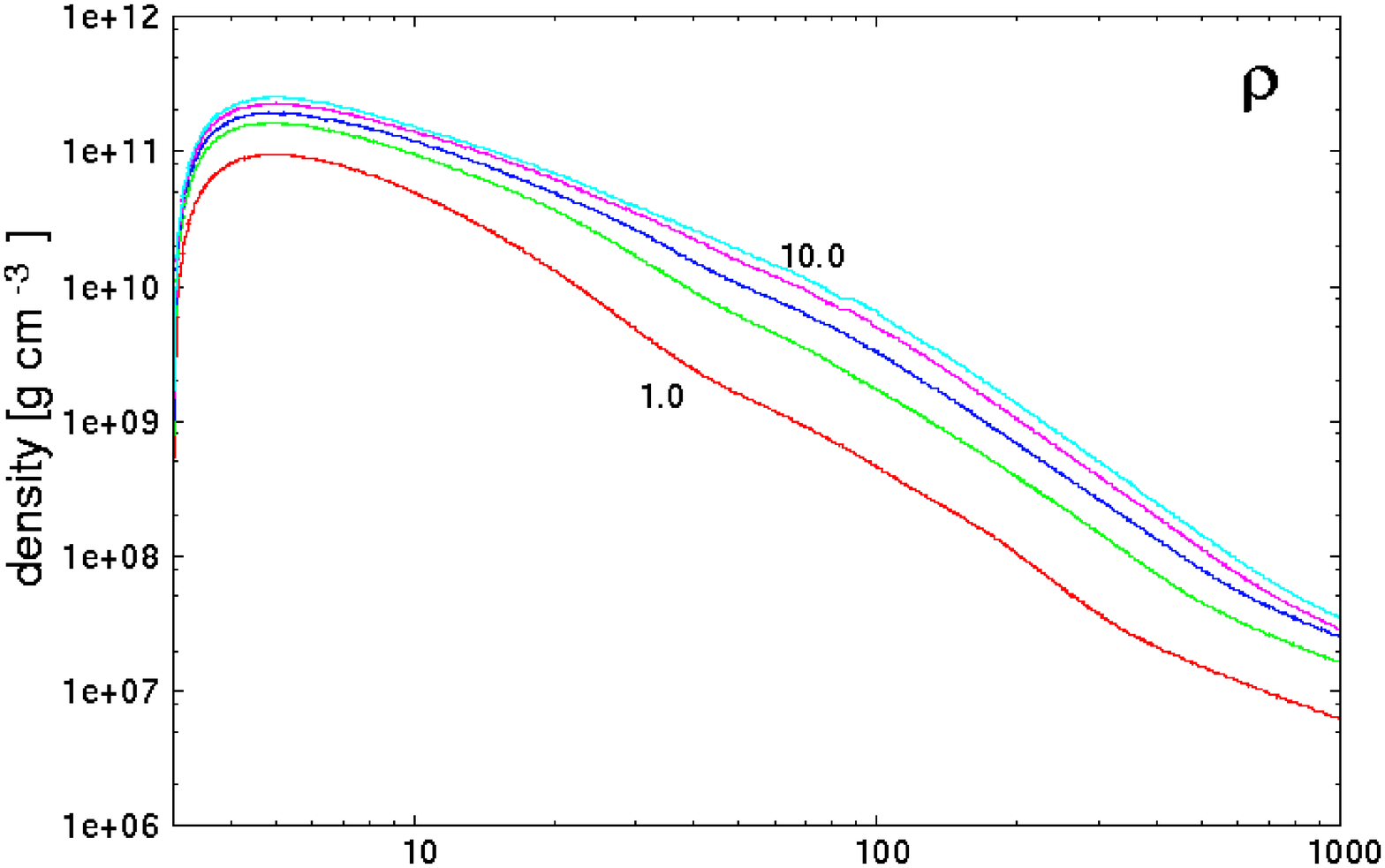}} &
      \resizebox{50mm}{!}{\includegraphics{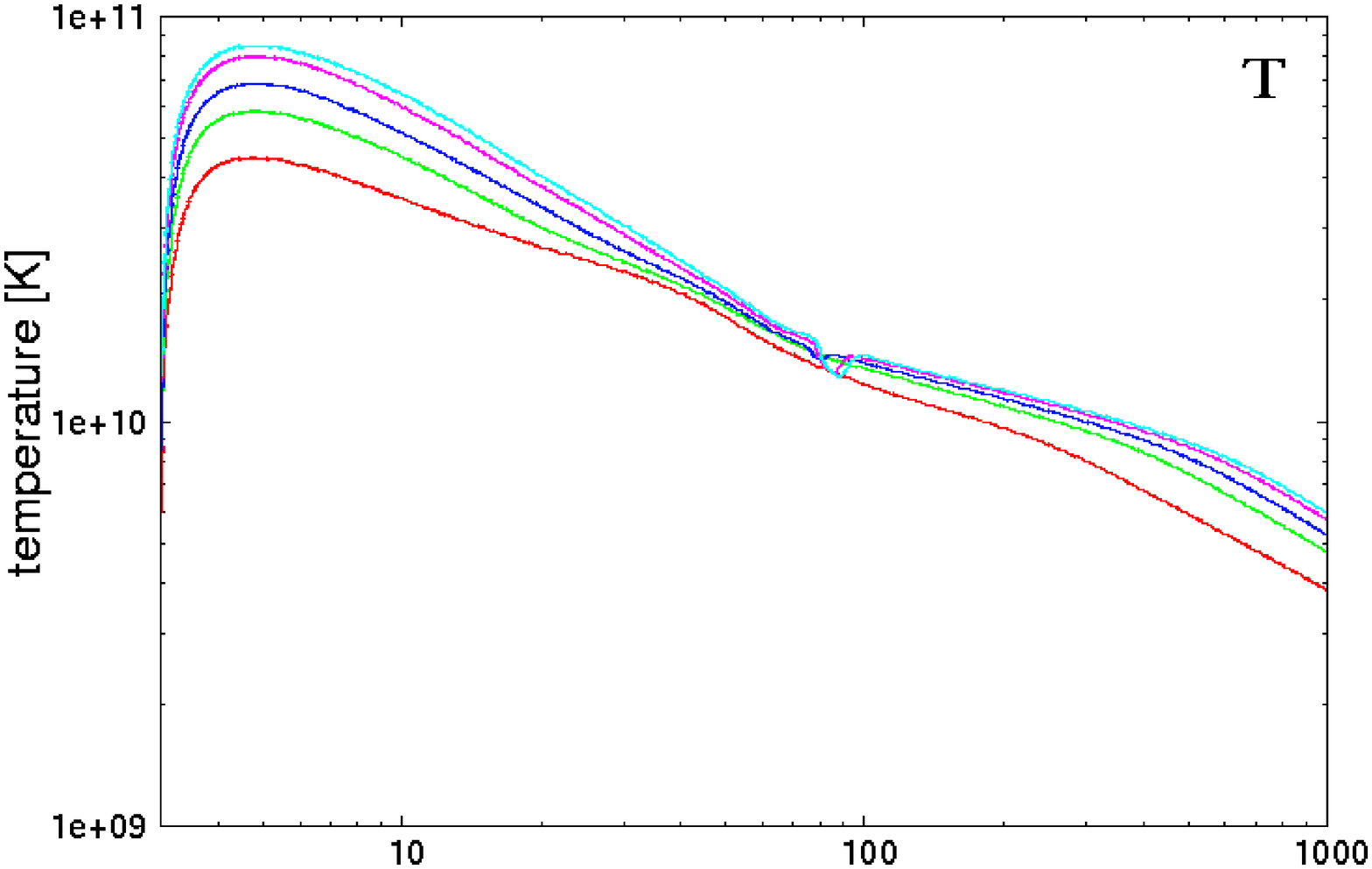}} &
      \resizebox{50mm}{!}{\includegraphics{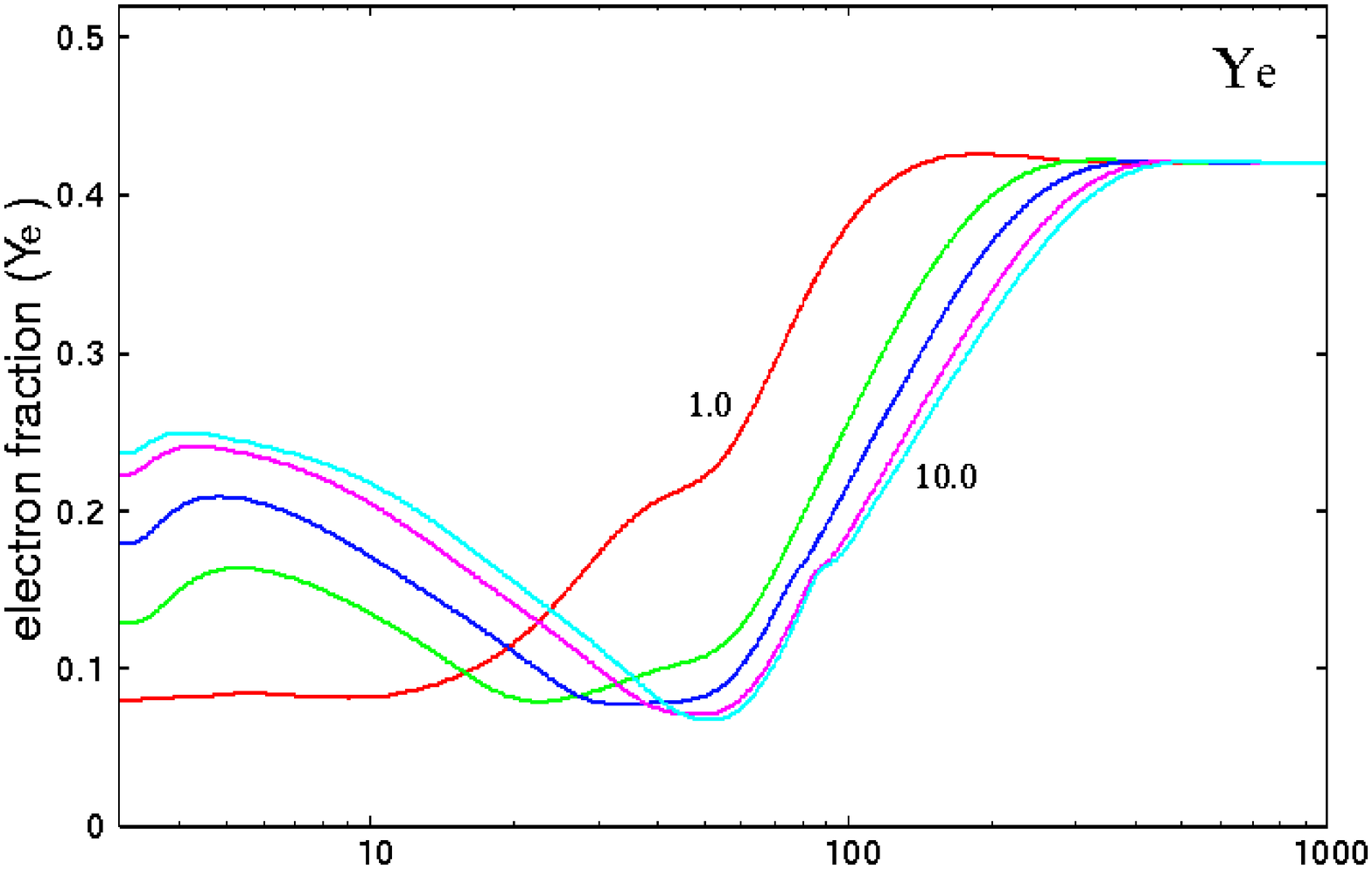}} \\
      \resizebox{50mm}{!}{\includegraphics{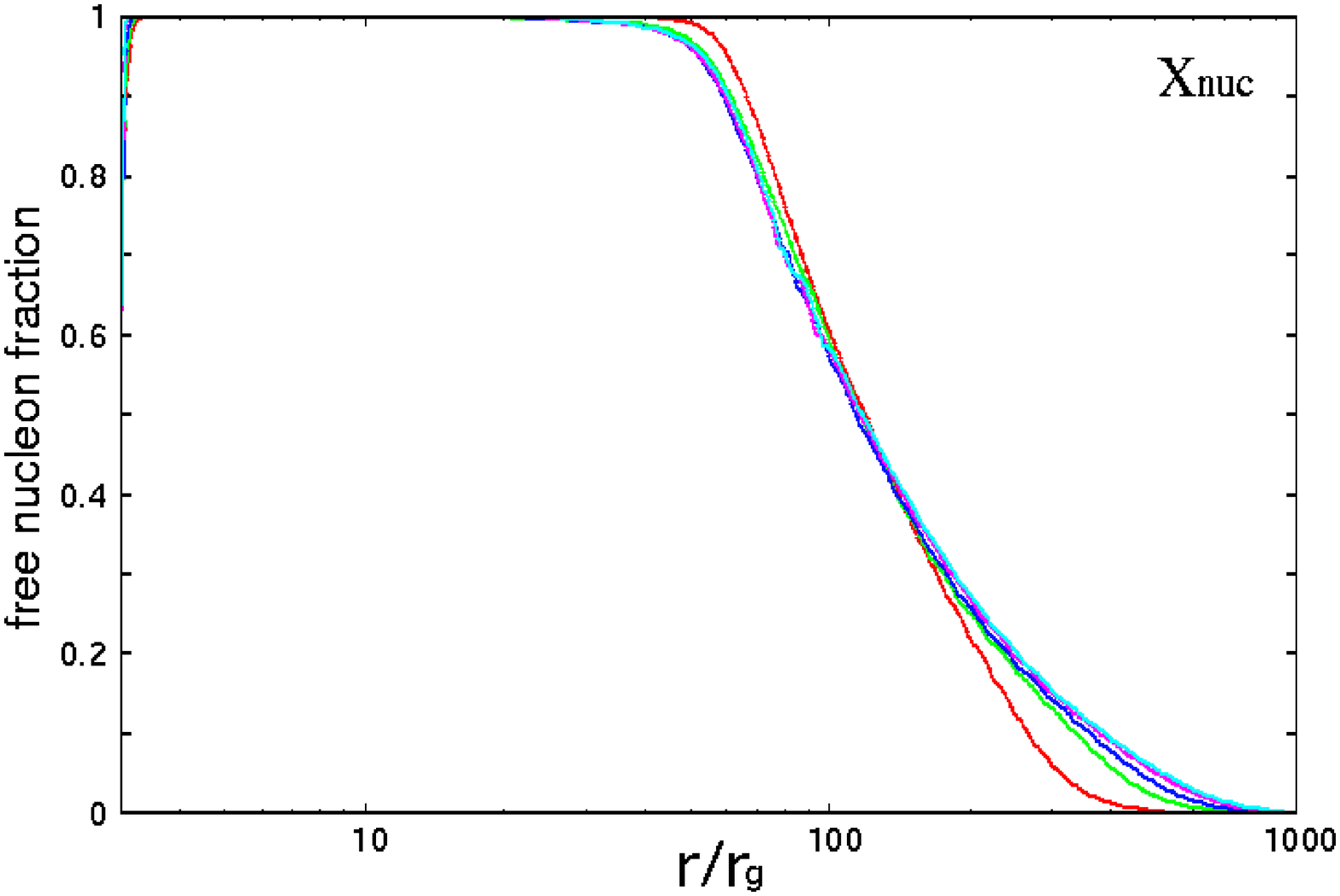}} &
      \resizebox{50mm}{!}{\includegraphics{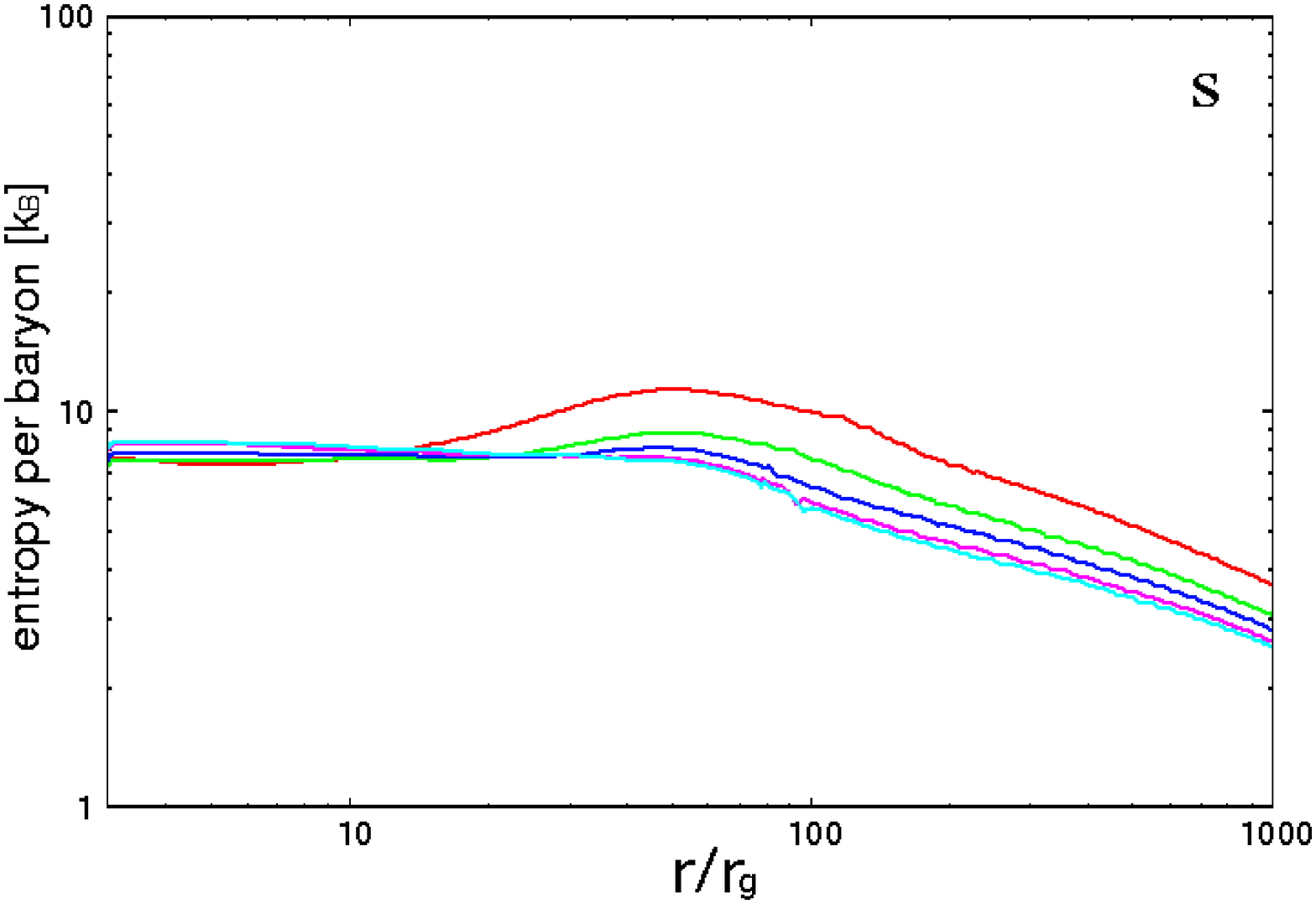}} &
      \resizebox{50mm}{!}{\includegraphics{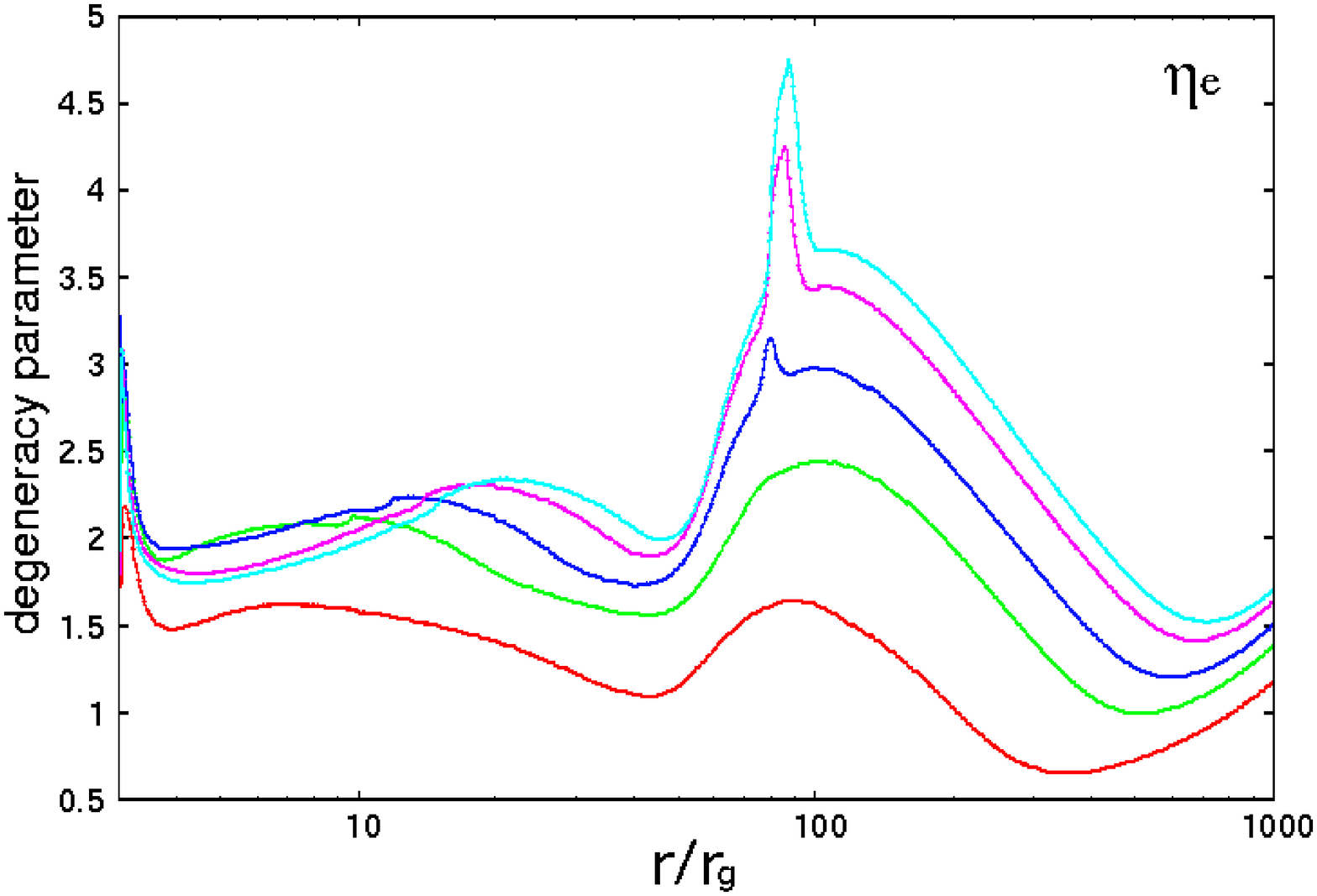}} \\
    \end{tabular}
    \caption{Same as Fig 1, but
 for the accretion flow with $\dot{M}=1.0$ ({\it red}), $3.0$ ({\it green}), $5.0$ ({\it blue}), $8.0$ ({\it purple}), and $10.0M_{\odot}~{\rm sec}^{-1}$
 ({\it cyan}).}
    \label{highacc}
  \end{center}
\end{figure}
\vfill
\clearpage
\vfill
\begin{figure}
\begin{center}
\resizebox{!}{10cm}{\includegraphics{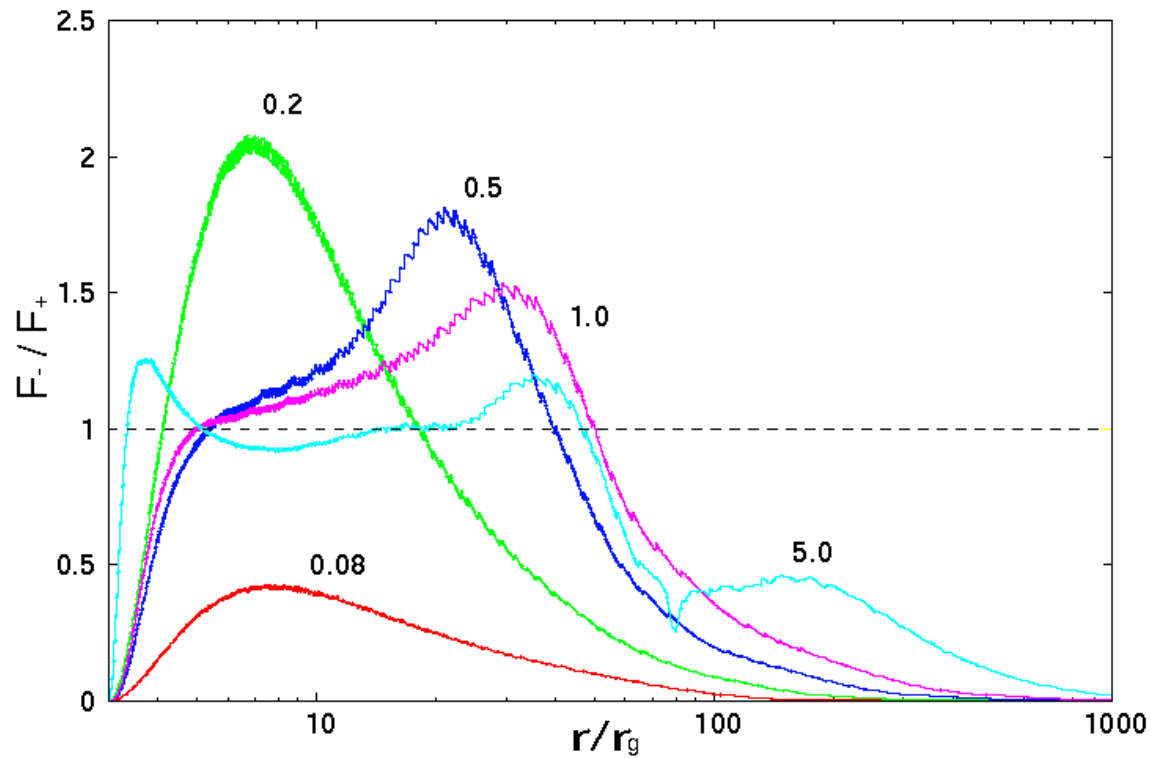}} \\
\caption{The ratios of neutrino cooling flux ($F_-$) to viscous heating rate per unit disk surface area ($F_+$) in NDAF with mass accretion rates
 of 0.08({\it red}), $0.2$ ({\it green}), $0.5$ ({\it blue}), $1.0$ ({\it purple}), and $5.0M_{\odot}~{\rm sec}^{-1}$ ({\it cyan}).}
\label{chratio}
\end{center}
\end{figure}
\vfill
\clearpage
\vfill
\begin{figure}
\begin{center}
\resizebox{!}{8cm}{\includegraphics{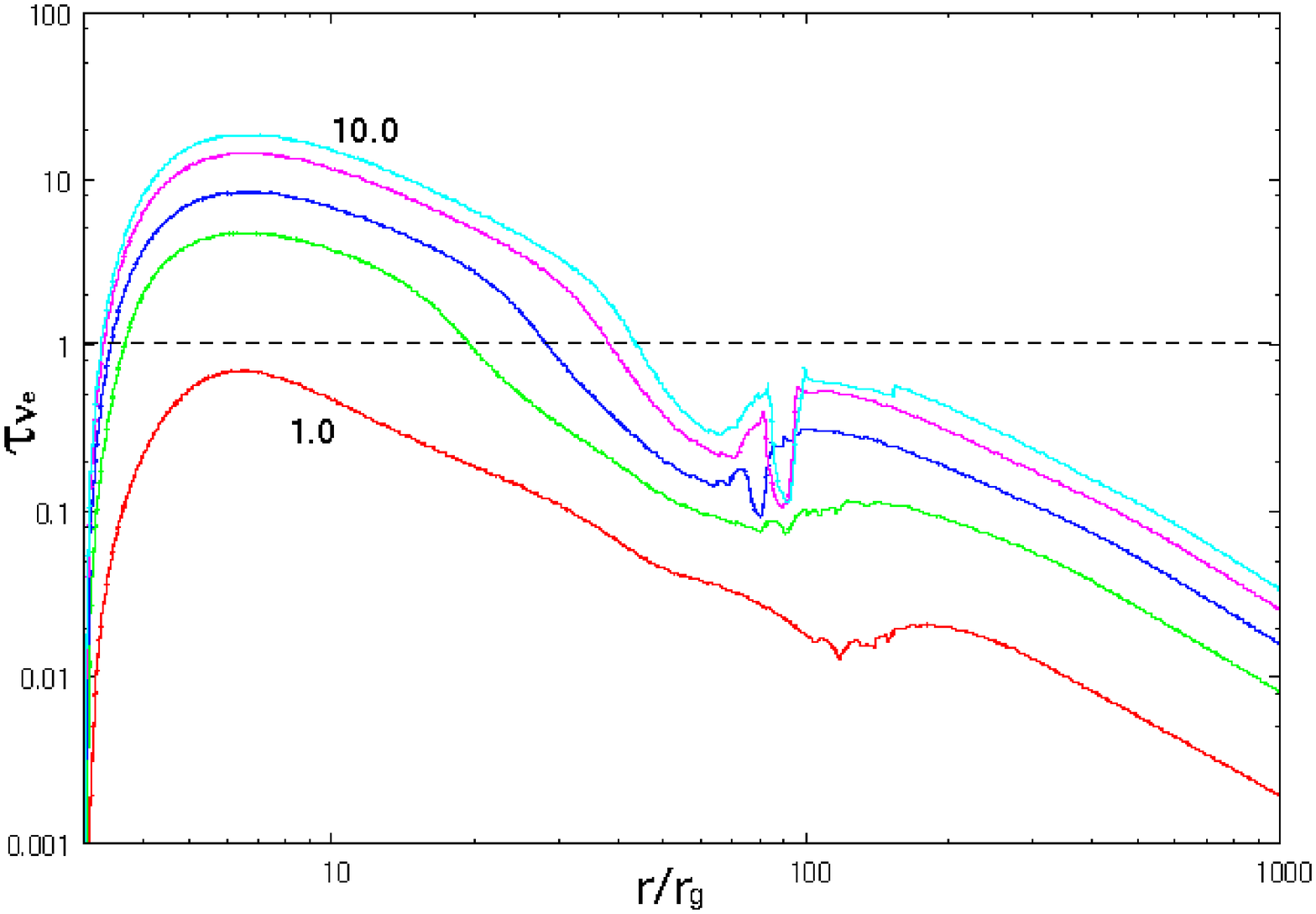}} \\
\resizebox{!}{8cm}{\includegraphics{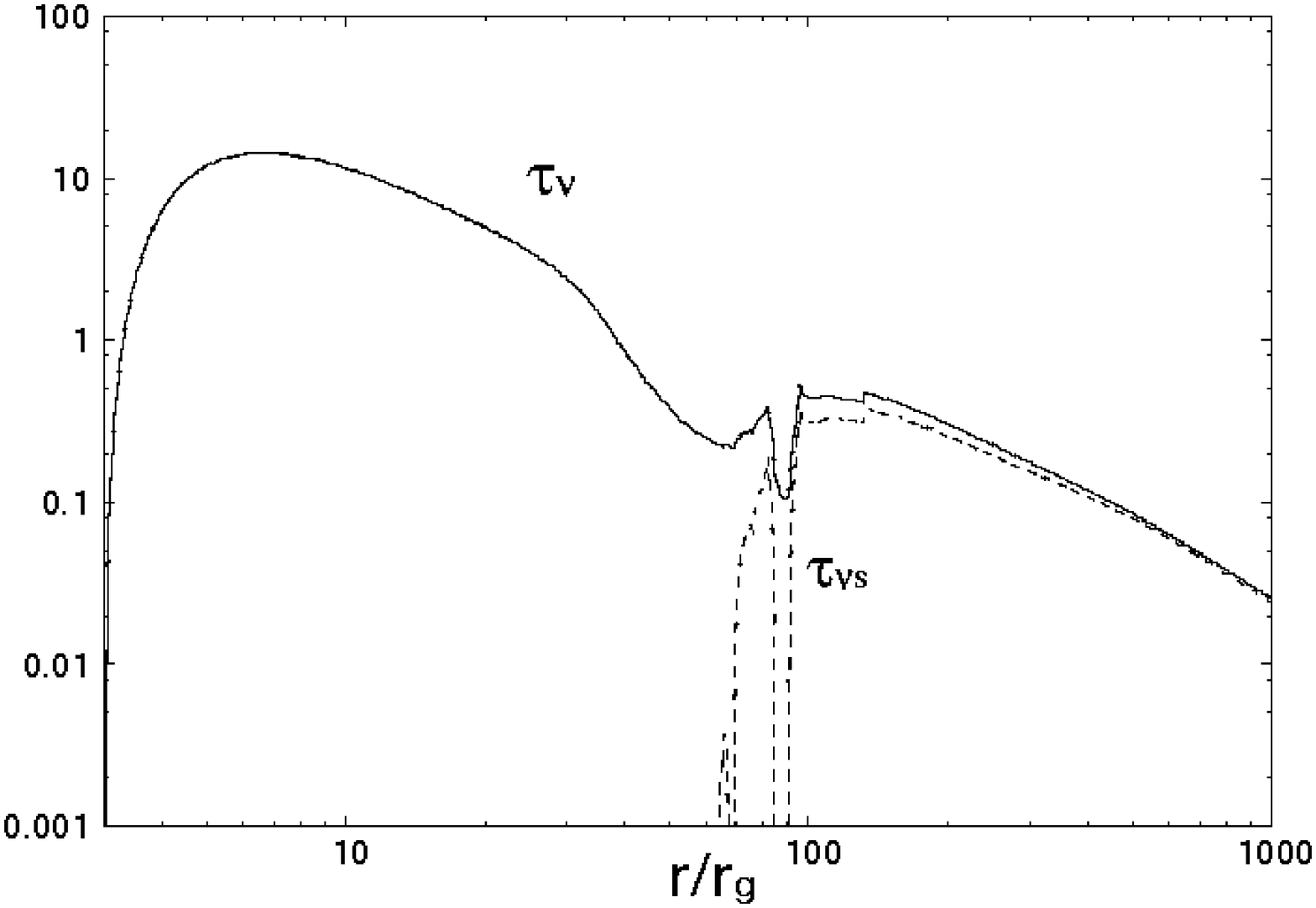}} \\
\caption{Upper panel: The radial profiles of the optical depth with respect to electron neutrinos in NDAF with mass accretion rates of 1.0({\it red}),
 $3.0$ ({\it green}), $5.0$ ({\it blue}), $8.0$ ({\it purple}), and $10.0M_{\odot}~{\rm sec}^{-1}$ ({\it cyan}).  Lower panel: The radial profiles
 of the optical depth ({\it solid})and the scattering optical depth by heavy nuclei ({\it dashed})with respect to electron neutrinos.}
\label{taue}
\end{center}
\end{figure}
\vfill
\clearpage
\vfill
\begin{figure}
  \begin{center}
    \begin{tabular}{cc}
      \resizebox{80mm}{!}{\includegraphics{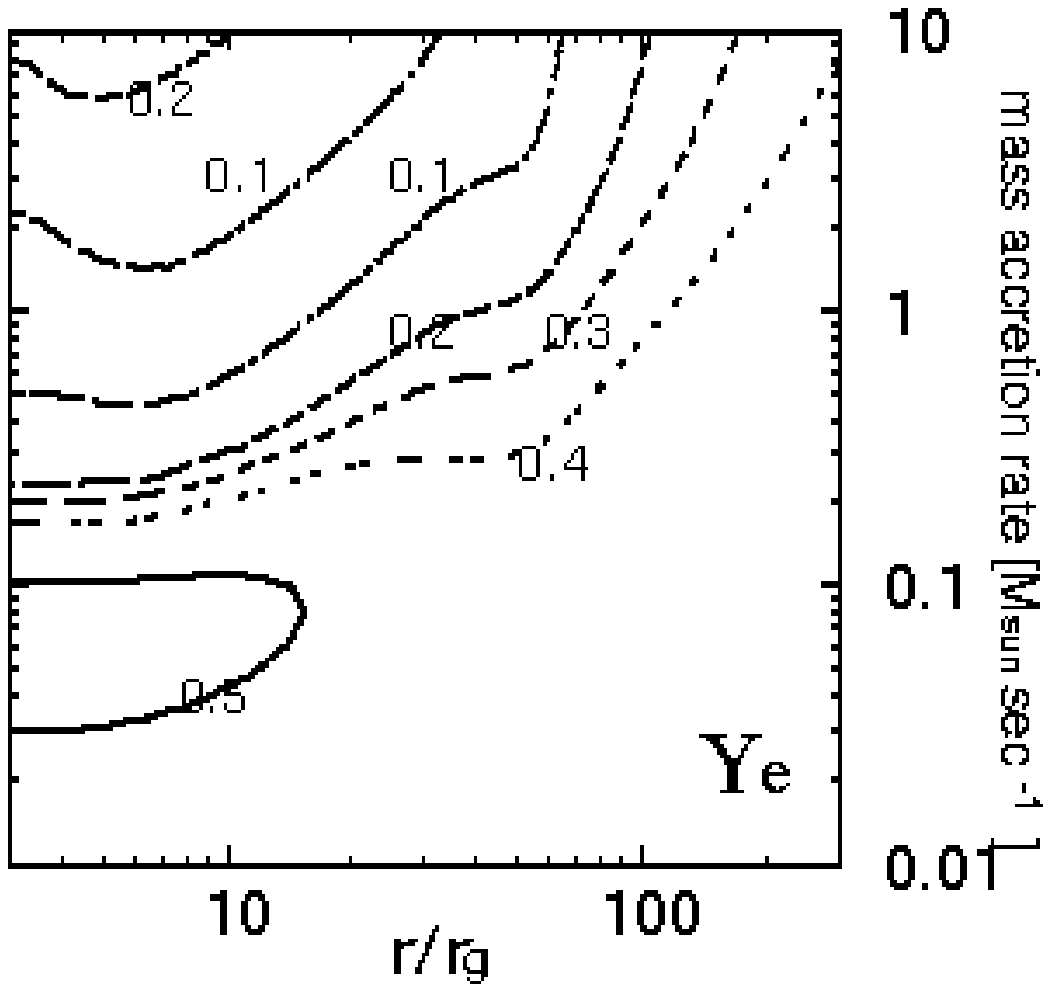}} &
      \resizebox{80mm}{!}{\includegraphics{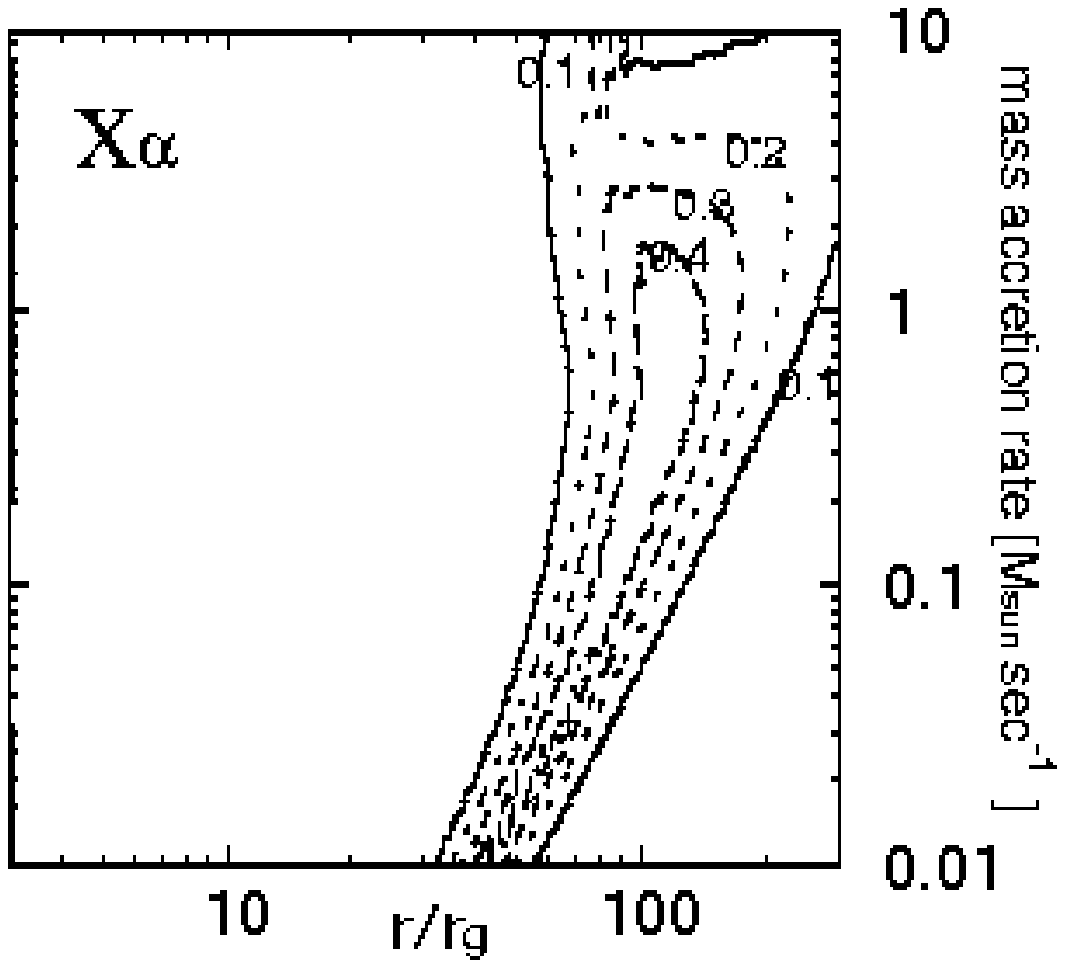}} \\
      \resizebox{80mm}{!}{\includegraphics{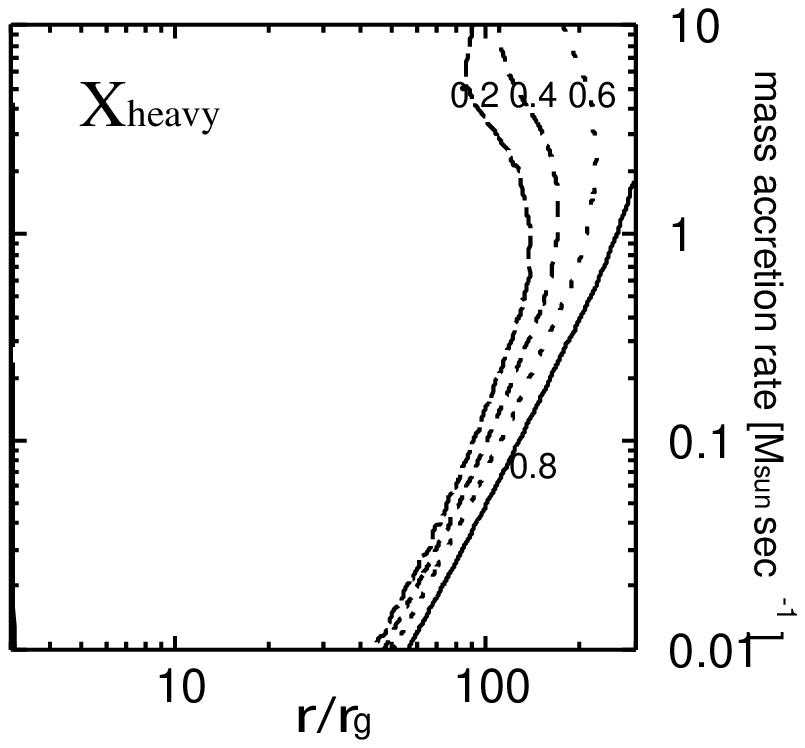}} &
      \\
    \end{tabular}
    \caption{Contours of electron fraction ({\it left-top}), $\alpha$-particle fraction ({\it right-top}), and heavy nucleus fraction ({\it left-bottom})
 on $(r, \dot{M})$ plane.}
    \label{contour}
  \end{center}
\end{figure}
\vfill
\clearpage
\vfill
\begin{figure}
\begin{center}
\begin{tabular}{cc}
\resizebox{80mm}{!}{\includegraphics{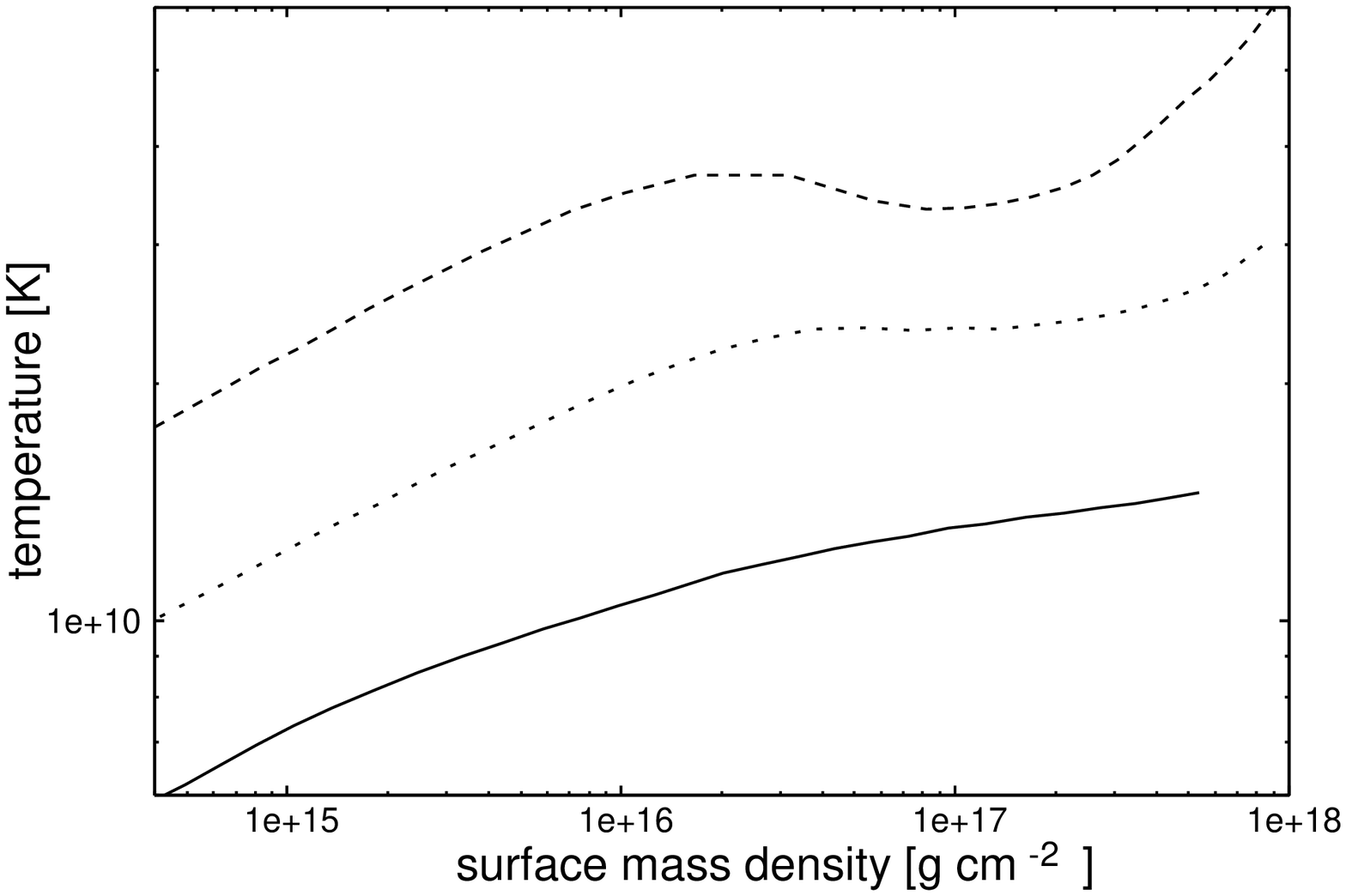}} &
\resizebox{80mm}{!}{\includegraphics{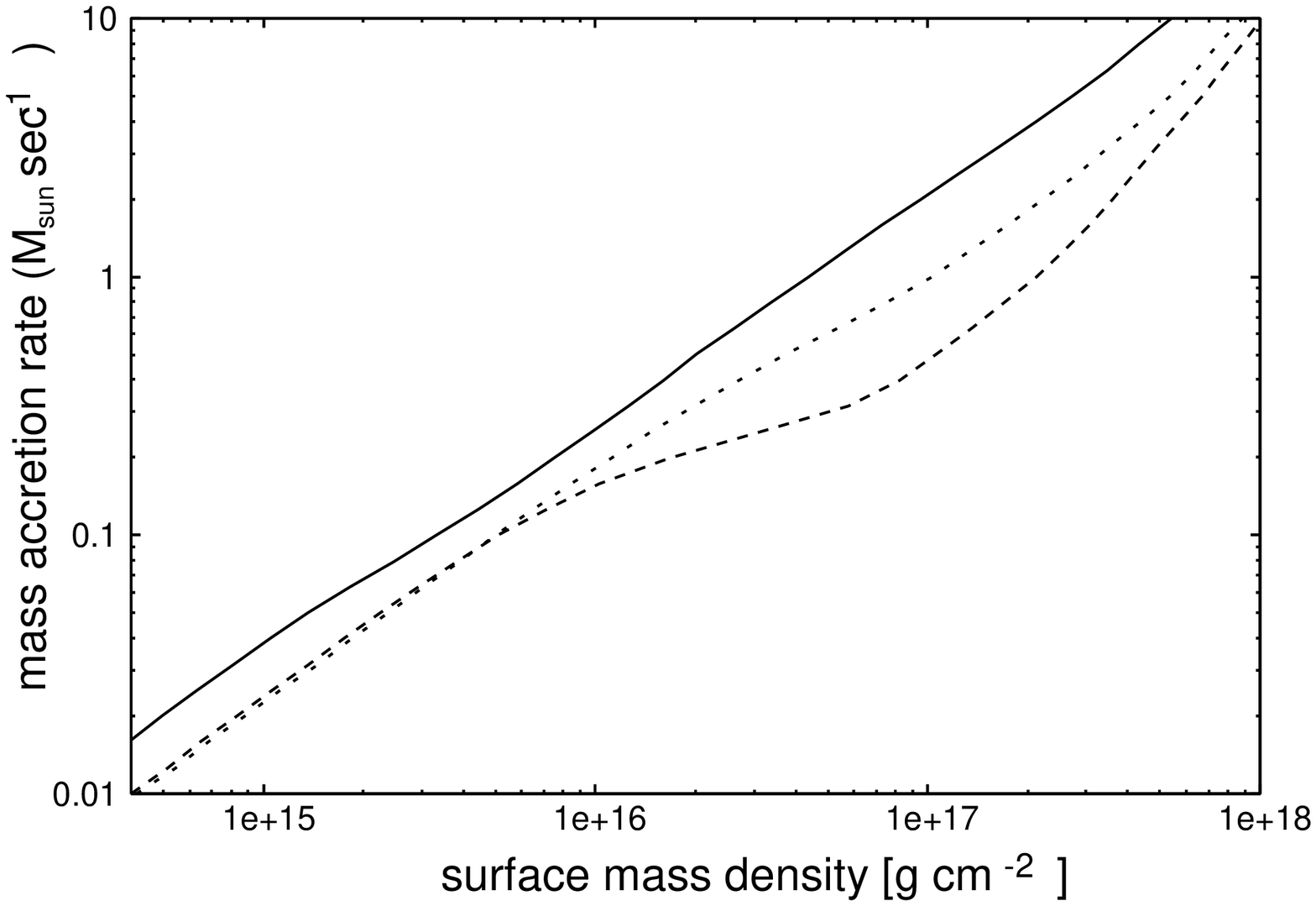}} \\
\end{tabular}
\caption{Thermal equilibrium curves on the $(\Sigma, T)$ plane and $(\Sigma, \dot{M})$ plane at radial distances of $r=100r_g$ ({\it red}), $30r_g$ ({\it blue}),
 and $10r_g$ ({\it green}).}
\label{eqcurve}
\end{center}
\end{figure}
\vfill
\clearpage
\begin{figure}
\begin{center}
\resizebox{!}{8cm}{\includegraphics{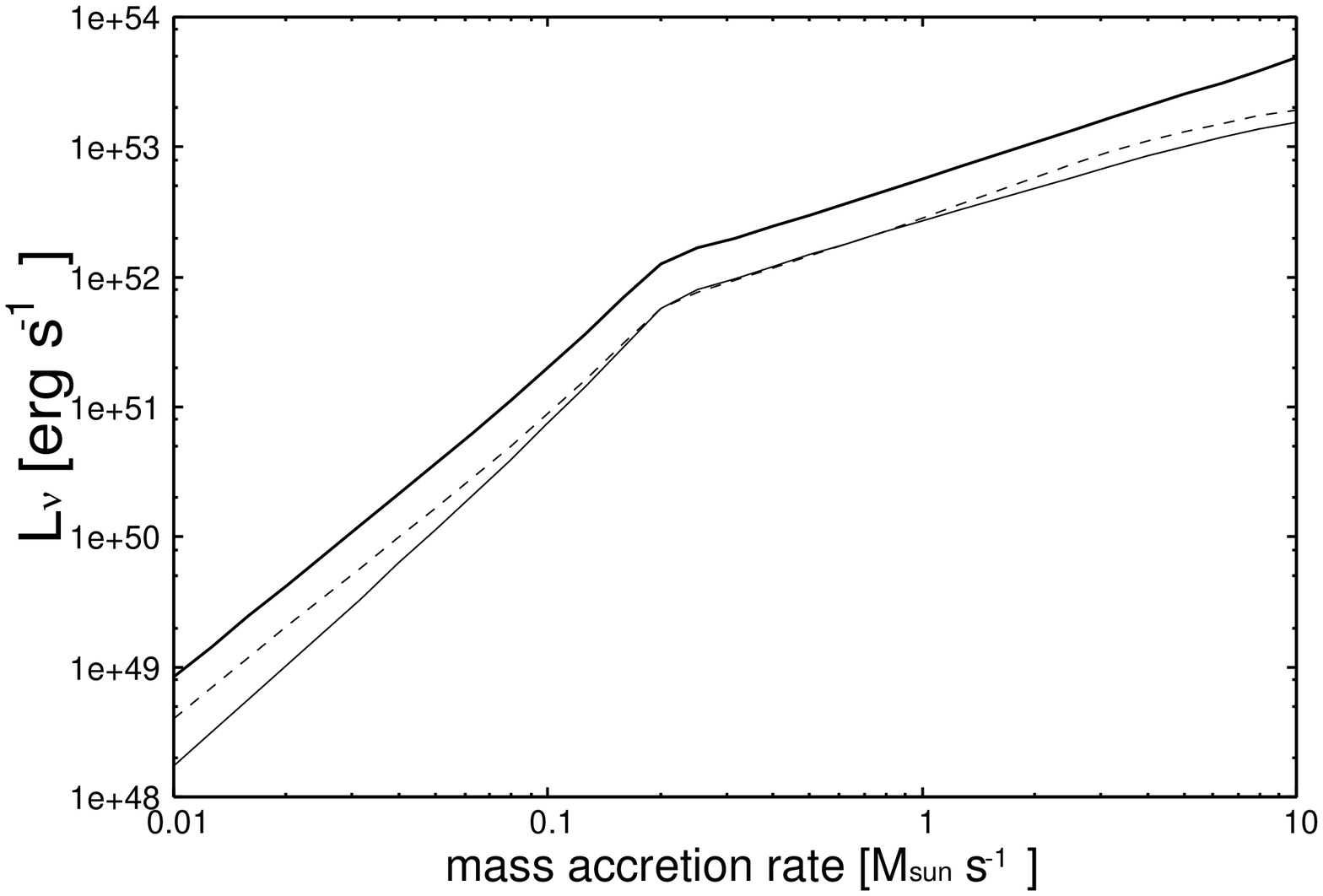}} \\
\resizebox{!}{8cm}{\includegraphics{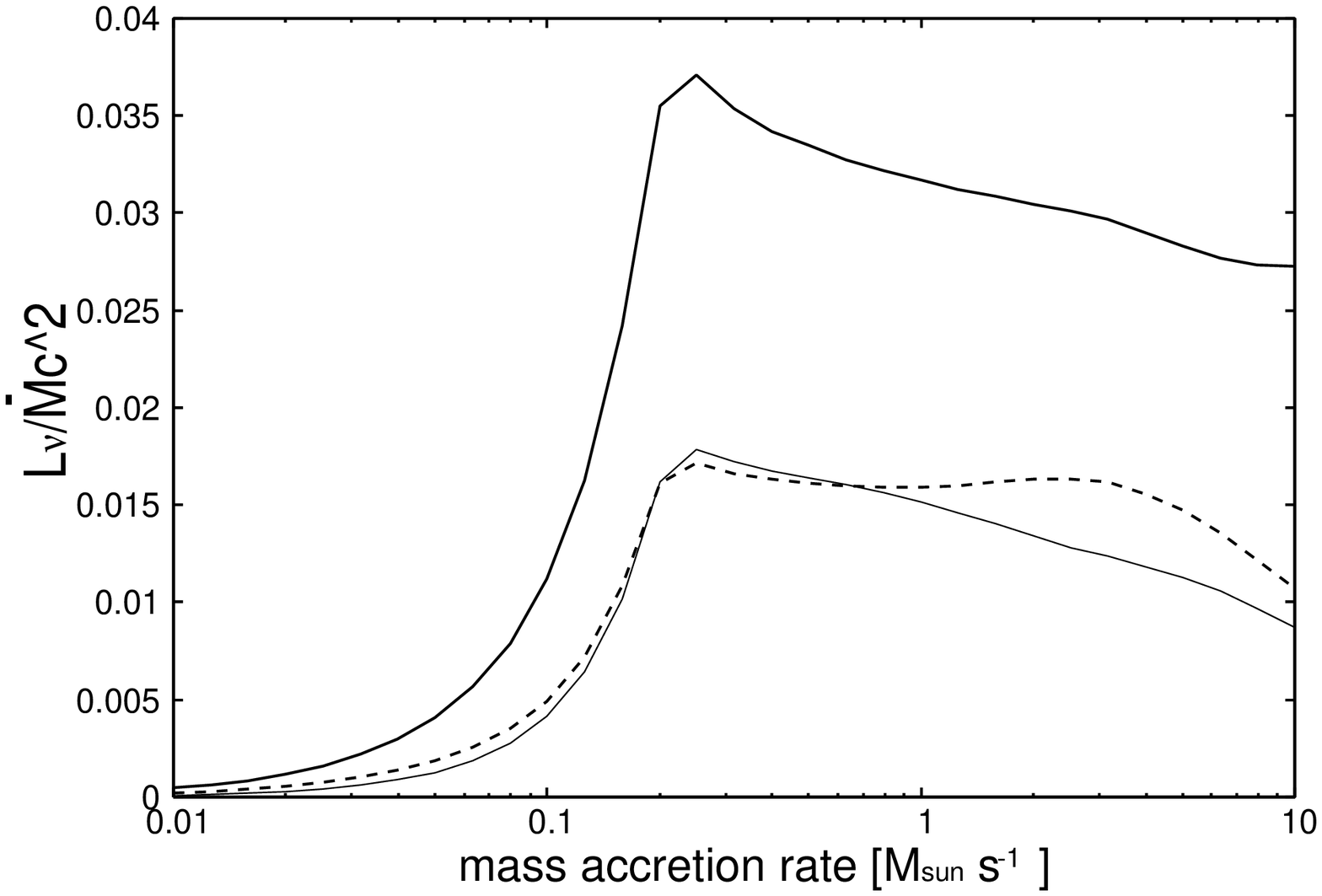}} \\
\caption{Upper panel: Total neutrino luminosity ({\it thick-solid}), $\nu_e$ luminosity ({\it thin-solid}), and $\bar{\nu}_e$ luminosity ({\it dashed}) from
 the hypercritical accretion disk as the function of mass accretion rate.  Lower panel:The ratio between total luminosity of neutrinos ({\it thick-solid}),
 $\nu_e$ luminosity ({\it thin-solid}), and $\bar{\nu}_e$ luminosity ({\it dashed}) from the whole surface of NDAF and total rest mass energy flowing into the central black hole per unit time ($\dot{M}c^2$) as the
 function of mass accretion rate.}
\label{eff}
\end{center}
\end{figure}
\end{document}